\documentclass[%
 reprint,
superscriptaddress,
 amsmath,amssymb,
 aps,
]{revtex4-2}

\usepackage{graphicx}
\usepackage{dcolumn}
\usepackage{bm}

\begin{document}


\title{Zero-field magnetic skyrmions in exchange-biased ferromagnetic-antiferromagnetic bilayers}

\author{M. Pankratova}
\affiliation{Department of Physics and Astronomy, Uppsala University, Box 516, SE-75120 Uppsala, Sweden}
\email[Corresponding author:\ ]{maryna.pankratova@physics.uu.se}

\author{O. Eriksson}
\affiliation{Department of Physics and Astronomy, Uppsala University, Box 516, SE-75120 Uppsala, Sweden}
\affiliation{Wallenberg Initiative Materials Science for Sustainability, Uppsala University, 75121 Uppsala, Sweden}

\author{A. Bergman}
\affiliation{Department of Physics and Astronomy, Uppsala University, Box 516, SE-75120 Uppsala, Sweden}

\date{\today}

\begin{abstract}
We report on the stabilization of ferromagnetic skyrmions in zero external magnetic fields, in exchange-biased systems composed of ferromagnetic-antiferromagnetic (FM-AFM) bilayers. By performing atomistic spin dynamics simulations, we study cases of compensated, uncompensated, and partly uncompensated  FM-AFM interfaces, and investigate the impact of important parameters such as temperature, inter-plane exchange interaction, Dzyaloshinskii-Moria interaction, and magnetic anisotropy on the skyrmions appearance and stability. The model with an uncompensated FM-AFM interface leads to the stabilization of individual skyrmions and skyrmion lattices in the FM layer, caused by the effective field from the AFM instead of an external magnetic field. Similarly, in the case of a fully compensated FM-AFM interface, we show that FM skyrmions can be stabilized. We also demonstrate that accounting for interface roughness leads to stabilization of skyrmions both in compensated and uncompensated interface. Moreover, in bilayers with a rough interface, skyrmions in the FM layer are observed for a wide range of exchange interaction values through the FM-AFM interface, and the chirality of the skyrmions depends critically on the exchange interaction. 
\end{abstract}

\maketitle


\section{\label{sec:intro}
Introduction}

Modern society currently produces an ever-increasing amount of data, leading to pressing challenges for data-storage technologies, one of which is storage density, i.e. storing bigger amount of data in smaller space. Moreover, data centers already use around one percent of global electricity demand, calling for less energy-hungry data storage technologies.  Magnetic skyrmions are nanoscale topological spin structures \cite{tokura2020magnetic, WANG2022169905,everschor2018perspective}, that are very promising candidates for data-carrying or for logic elements in developing spintronic devices \cite{everschor2018perspective}. Skyrmions are stable due to the combination of Heisenberg exchange interactions, asymmetric exchange interactions, so-called Dzyaloshinskii-Moriya (DMI) interaction, and an external magnetic field. DMI, in contrast with the Heisenberg exchange interaction, tends to favour perpendicular magnetic moment orientations and introduces chirality to the system, which is central for the creation of skyrmions.   

Skyrmions are often stabilized using external magnetic fields \cite{everschor2018perspective}, however, for full exploration of spintronic devices involving skyrmions it is desirable to stabilize skyrmions in the absence of a magnetic field. Many works have been devoted to skyrmion stabilization in zero applied field, e.g. in bilayers and multilayers \cite{ajejas2023densely,JIANG20171, chen2023tailoring,rana2021imprint,mallick2022current, brandao2019observation}. It was shown that the DMI interaction can be transferred from an antiferromagnet (AFM) to a ferromagnet (FM) in FM-AFM bilayers to observe skyrmions. Recently, AFM skyrmions in zero magnetic were predicted in FM-AFM bilayers on triangular lattice by Monte Carlo simulations \cite{MOHYLNA2022128350}. 

Moreover, it was experimentally shown that skyrmions can be stabilized in zero external magnetic field by using the exchange bias phenomena (EBP) in ferromagnets \cite{yu2018room,rana2020room, dieny2024comparison} and antiferromagnets \cite{rana2021imprint}. EBP \cite{PhysRev.105.904} manifest itself in the shift of magnetic hysteresis along the field axis, and appears in systems with contacting FM-AFM, and other systems. The simplest model of exchange bias involves an uncompensated antiferromagnetic interface \cite{pankratova2017magnetization,grechnev2012influence} (non-zero magnetic moment of the AFM interface layer), therefore, an effective field acts from the AFM to the FM layer. The resulting effective field can be used to stabilize magnetic skyrmions in a FM, instead of an external applied magnetic field. However, this uncompensated model is known to lead to an overestimation of the shift of the hysteresis loop, when compared with experimental studies. Therefore, other models assumed a compensated FM-AFM interface (both AFM sublattices are presented at the interface and couple with equal strength to the FM)\cite{kiwi1999exchange}. Koon \cite{PhysRevLett.78.4865,stamps2000mechanisms} suggested that due to the competition between couplings in AFM with a FM, the interface spins could deviate from the easy axis, and form canted states, leading to the small effective field observed in experiments, that is manifested by EBP with rather small shifts of the hysteresis curve. 

Another important factor to consider, while studying EBP, is the interface roughness \cite{stamps2000mechanisms}. In most experiments some roughness or geometrical defects are present at the interface. Also, it was shown that exchange interaction through the interface is strongly impacted by the interface roughness \cite{lu2000effect}.  It was demonstrated experimentally in Ref.\cite{kappenberger2003direct} that only about 7$\%$ of the spins at the interface are uncompensated and contribute to the EBP. It was confirmed theoretically in Ref.\cite{kovalev2014field,pankratova2015model} that only several percent of uncompensated moments are enough to induce EBP. Also, it was shown that roughness and defects impacts interfacial DMI \cite{carvalho2023correlation} and therefore skyrmion stability. In addition, roughness has a strong impact on skyrmion dynamics, and results for rough and perfect interface cases can significantly differ \cite{roughskyrm}, which makes accounting of interface defects and roughness vital.

In this work, we study the possibility of stabilizing skyrmions in zero external magnetic fields using the exchange bias phenomena. We investigate various types of FM-AFM interfaces and the range of parameters, such as exchange interaction, DMI,  magnetic anisotropy, and temperature allowing skyrmion stabilization. The paper is organized as follows: we start with describing details of atomistic spin dynamics simulations, then we investigate the case of an perfect interface. Finally, we study the impact of disorder or (geometrical) roughness of the interface on the magnetic structure of FM and AFM films.

\section{\label{sec:simu} Atomistic spin dynamics simulations}

Atomistic spin dynamics (ASD) is governed by Landau-Lifshits-Gilbert equation:
\begin{align}\label{eq1}
\frac{d \boldsymbol{m}_i}{d t}
& = - \frac{\gamma}{(1+\alpha ^2)}\boldsymbol{m}_i\times(\boldsymbol{B}_i+\boldsymbol{B}_i^{\mathrm{fl}})\\
& \quad - \frac{\gamma}{(1+\alpha ^2)}\frac{\alpha}{m_i}\boldsymbol{m}_i\times(\boldsymbol{m}_i \times [\boldsymbol{B}_i+\boldsymbol{B}_i^{\mathrm{fl}}]),\notag
\end{align}
where $\boldsymbol{m_i}$ represents an atomic magnetic moment, $m_i$ and  $\gamma$ are the saturation magnetization and the gyromagnetic ratio correspondingly. We obtain an effective exchange field $\boldsymbol{B}_i = - \partial H_{\mathrm{SD}}/\partial \boldsymbol{m}_i$ from the spin Hamiltonian, $H_{SD}$. The Hamiltonian used in this work includes FM, AFM parts and interaction between FM and AFM subsystems: 
\begin{equation}
    H_{\mathrm{SD}}= H_{\mathrm{FM}}+H_{\mathrm{AFM}}+H_{\mathrm{FM/AFM}}.
\end{equation}

Concretely:
\begin{align}\label{eq:H_SD}
H_{\mathrm{SD}} & = - \frac{1}{2} \sum_{ij} J_{ij}^{\mathrm{FM}} \boldsymbol{m}_i \cdot \boldsymbol{m}_j - \frac{1}{2} \sum_{kl} J_{kl}^{\mathrm{AFM}} \boldsymbol{m}_k \cdot \boldsymbol{m}_l \\
& \quad
- \frac{1}{2} \sum_{ik} J_{ik}^{\mathrm{FM/AFM}} \boldsymbol{m}_i \cdot \boldsymbol{m}_k
- \sum_{ij} D_{ij}^{\mathrm{FM}} \cdot( \boldsymbol{m}_i \times \boldsymbol{m}_j) \notag \\
& \quad 
- H_{\mathrm{ext}} \Bigl( \sum_{i} m_i^z + \sum_{k} m^z_k \Bigr) - K^{\mathrm{AFM}} \sum_{k} (m^z_k)^2,\notag 
\end{align}
where $J_{ij}$,  $J_{kl}$, and  $J_{ik}$ are the exchange interaction in FM, AFM, and through FM-AFM interface correspondingly. $H_{ext}$ denotes external magnetic field. We take into account DMI interaction $D_{ij}^{\mathrm{FM}}$ and magnetocrystalline anisotropy $K^{\mathrm{AFM}}$ for AFM. The anisotropy term is included mainly to ensure that the AFM layers stay pinned during the simulations, and is thus larger than what is normally expected in real systems.

In these Langevin-type simulations we employ stochastic fields, $\boldsymbol{B}_i^{\mathrm{fl}}$ as white noise with properties $\langle B_{i,\mu}^{\mathrm{fl}}(t) B_{j,\nu}^{\mathrm{fl}}(t') \rangle=2D_M \delta_{ij}\delta_{\mu\nu}\delta(t-t')$. In our simulations, we use $D_M= \alpha k_B T/\gamma m$, $D_L= \nu M k_B T$ , where $T$ and $k_B$ are temperature and Boltzmann constant respectively (for details see e.g. Ref.\,\cite{eriksson2017atomistic}).

The atomistic spin dynamics simulations are complemented by Monte Carlo (MC) calculations on the spin Hamiltonian. Since both LLG and MC simulations are based on the same Hamiltonian, they exhibit the same energetics and give the same ground state. However, the ground state search for the simulated systems is often complicated by the large energy barriers present in the systems due to the inclusion of strong DMI and anisotropies. To improve the efficiency of the simulations we therefore consider several different starting states for our simulations, including random, ferromagnetic, and spin-spiral states.

In our simulations we use the spin simulation package UppASD \cite{eriksson2017atomistic,UppASD} and  study the FM-AFM bilayer with square lattice consisting of three atomic layers for AFM and one atomic layer for FM. The simulation cell size vary between models, due to the reasons given above, and it is between $120 \times 120 \times 4$ and $160 \times 160 \times 4$. The value of magnetic moment in all simulations is the same for FM and AFM, and equals 1 $\mu_b$/atom. The value of exchange interactions in FM and AFM is fixed in all simulations and the ratio is  $|J^{\mathrm{AFM}}/J^{\mathrm{FM}}|=1$.
Most simulations are performed at zero temperatures, unless stated otherwise.

The topological charge $Q_{i}^{ab}$ is calculated using the method proposed in \cite{PhysRevB.39.7212}, since the model is atomistic and the expression for the continuum limit cannot be used.
In particular, each unit cube of the lattice is divided into 12 triangles, two for each face. Then the chirality enclosed by the unit cube is given by the expression:

\begin{equation}\label{eq:Nsk}
Q_{\mathrm{i}}^{ab} = \sum_{i=1}^{12} \boldsymbol{m_i} \cdot [\boldsymbol{m_a}\times \boldsymbol{m_b}],
\end{equation}

\noindent
where $\boldsymbol{m_i} \cdot [\boldsymbol{m_a}\times \boldsymbol{m_b}]$ is the chirality of the triangle formed by three neighboring spins.
Spins are taken in in counter-clockwise direction according to \cite{PhysRevB.39.7212}.

\section{\label{sec:res} Results}

\subsection{Uncompensated FM-AFM interface}

\begin{figure}[h!]
 \begin{tabular}{ccc}
(a) & (b) & (c) \\
\includegraphics[width=0.15\textwidth]{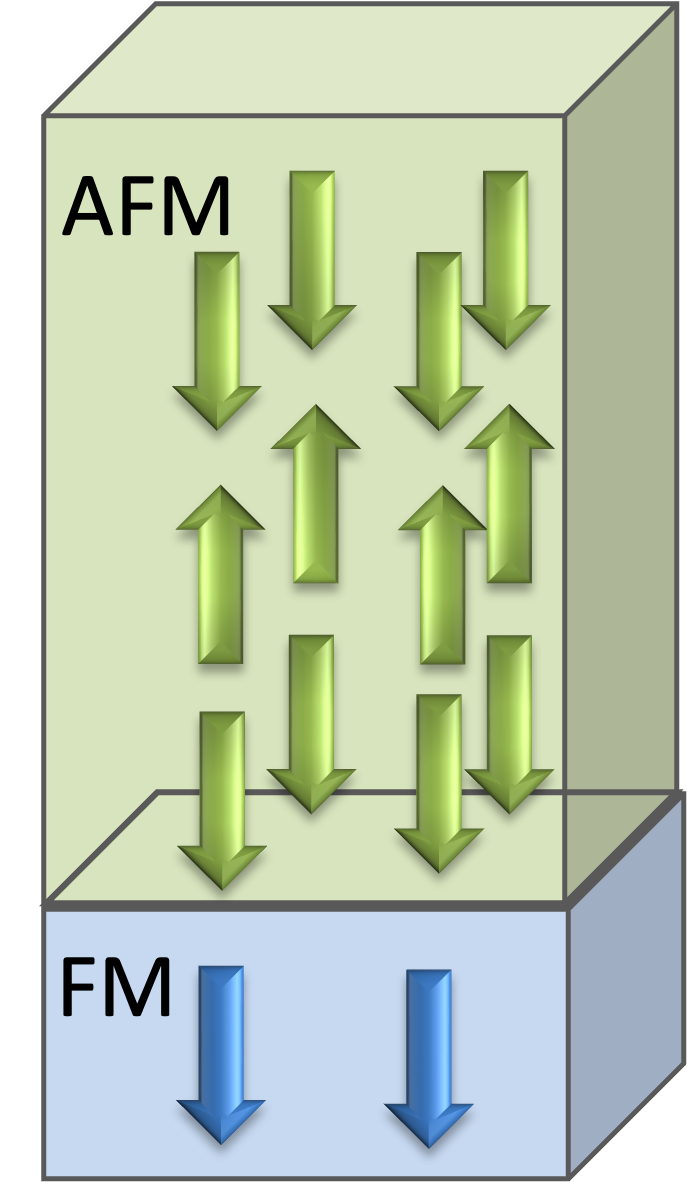}  & 
\includegraphics[width=0.15\textwidth]{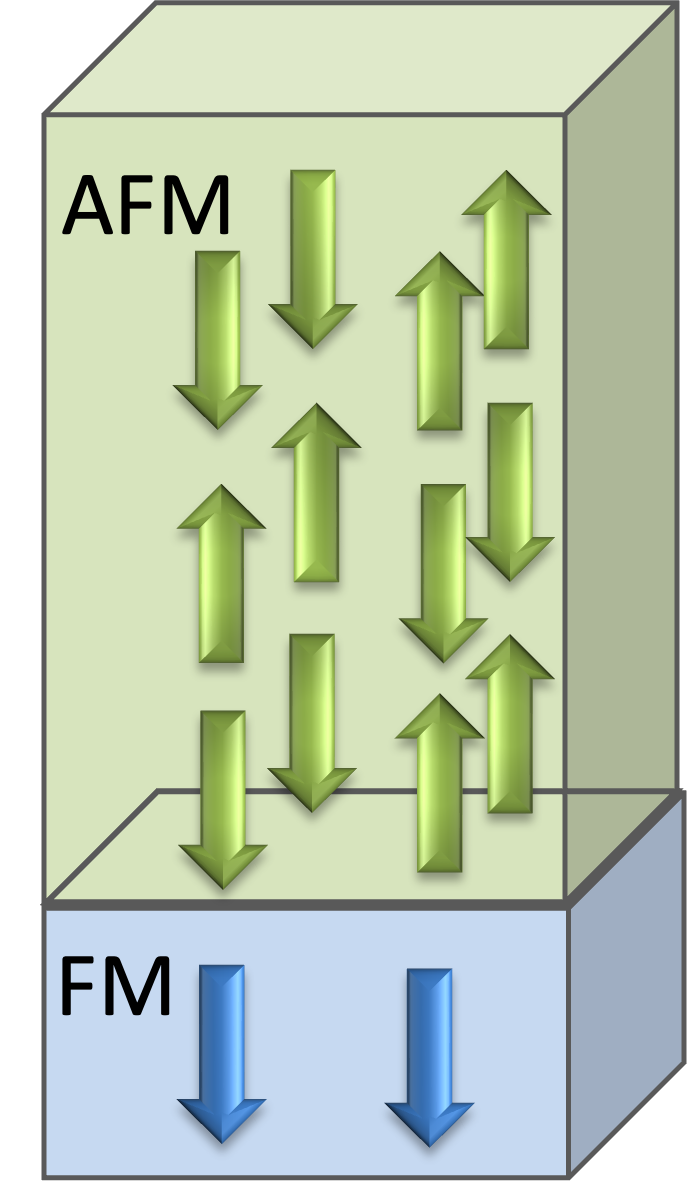}& 
\includegraphics[width=0.15\textwidth]{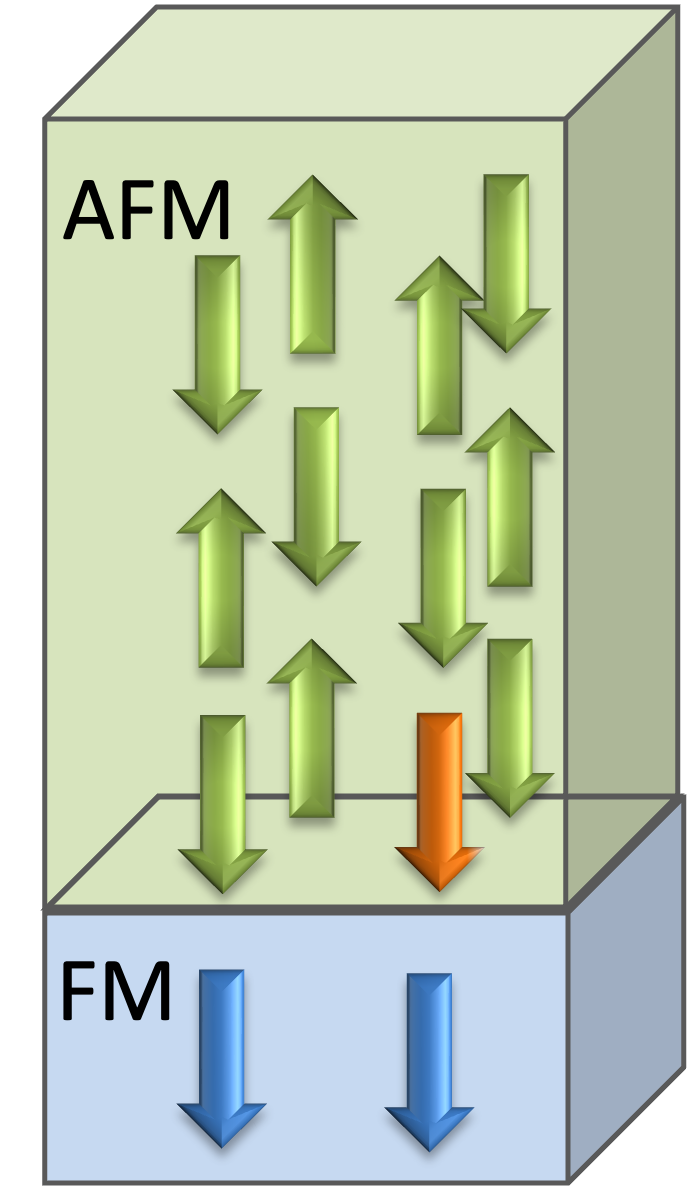}
\end{tabular}
\vspace{-0.2cm}
\caption{The scheme of uncompensated (a), compensated (b), and partly-uncompensated (c) AFM interface models. In all our calculations we use one FM layer and three AFM layers.}
\vspace{-0.2cm}
\label{fig:model_uncom_com}

\end{figure}

\begin{figure}[h!]
\centering
\includegraphics[width=0.5\textwidth]{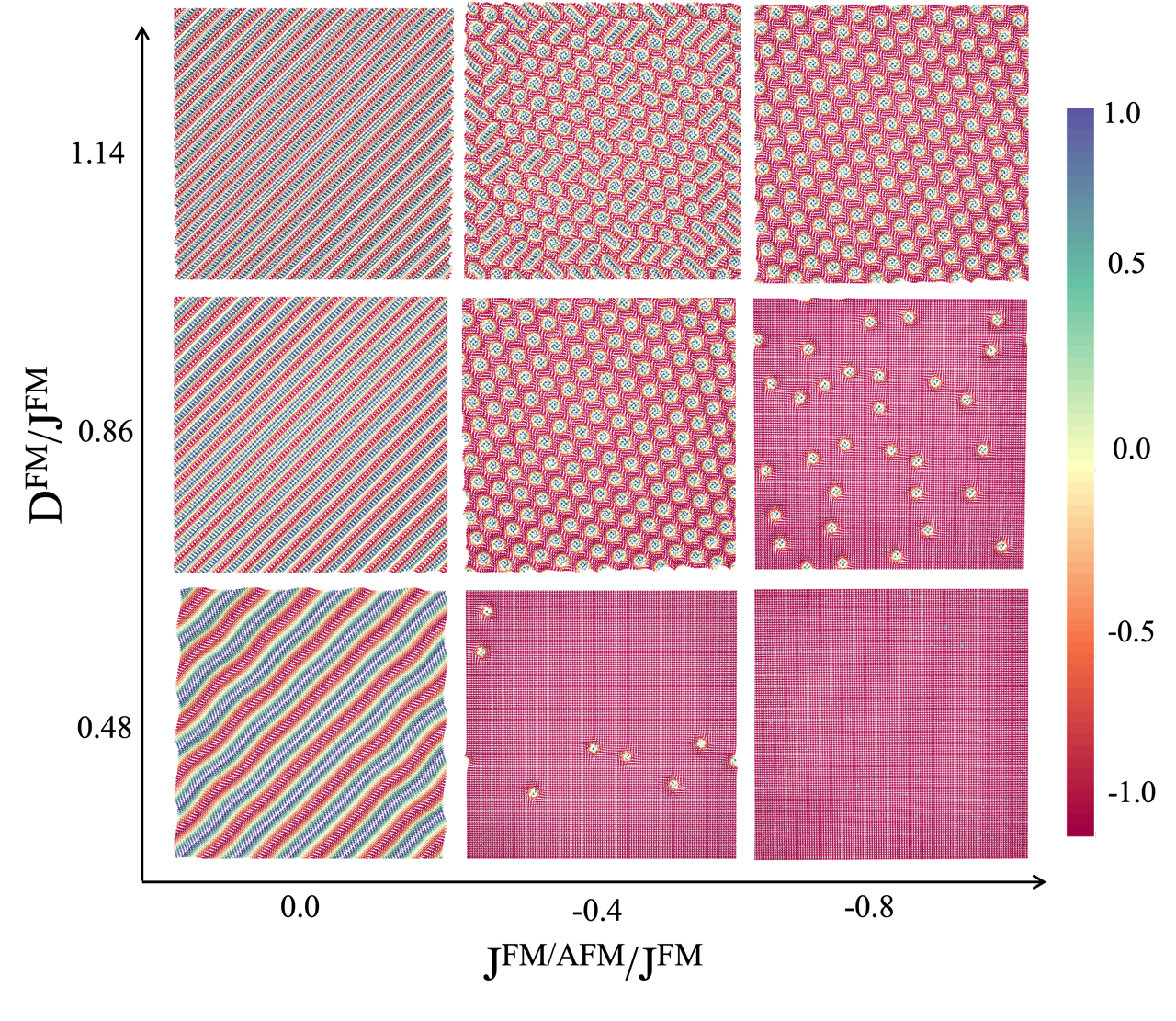}
\vspace{-0.2cm}
\caption{Ferromagnetic skyrmions in FM/AFM bilayer in zero external magnetic field for various values of exchange interaction through the FM/AFM interface and the strength of the DMI for interaction. Note that in the figure these two interactions are divided by the strength of the exchange interaction of the ferromagnetic layer. Magnetic anisotropy use din calculations is $K^{\mathrm{AFM}}/J^{\mathrm{FM}}=-0.5$.
}
\vspace{-0.2cm}
\label{fig:skyrm_uncomp}
\end{figure}

We start with an uncompensated model of the AFM interface. This means that the magnetic moments in each AFM layer are parallel to each other with $J^{\mathrm{AFM}}_{\mathrm{inlayer}} > 0$ with an antiparallel coupling between layers, $J^{\mathrm{AFM}}_{\mathrm{interlayer}} < 0$  (the so-called A-type AFM). Therefore, the total magnetic moment of the AFM interface layer in contact to the FM layer is non-zero and there is an effective exchange field that acts on FM film (see Fig.\ref{fig:model_uncom_com}a). 
Moreover, we assume a sizeable magnetic anisotropy $K^{\mathrm{AFM}}$ in each AFM layer, so AFM magnetic moments are almost "frozen" at their easy axis positions during the reversal  of an external magnetic field \cite{grechnev2012influence,pankratova2017magnetization}. Due to the effective field acting on the FM layer, we observe an exchange bias phenomenon with the shift of the magnetic hysteresis depending on the strength of exchange interaction $J^{\mathrm{FM/AFM}}$ through the FM-AFM interface and on the magnetic anisotropy in the AFM layer. The appearance and stabilization of skyrmions will also depend on the strengths of this effective field, and therefore, the exchange coupling $J^{\mathrm{FM/AFM}}$. In Fig.\ref{fig:skyrm_uncomp}  we show the magnetic structure of the FM layer for various values of DMI, and exchange interaction through the FM-AFM interface $J^{\mathrm{FM/AFM}}$ from simulations without an external magnetic field. One can see that in the absence of exchange interaction through the FM-AFM interface we observe stripe domains in FM (Fig.\ref{fig:skyrm_uncomp}, first column). However, when the exchange interaction, $J^{\mathrm{FM/AFM}}$, is increased, single skyrmions, then a skyrmion lattice, appears, as can be seen in Fig.\ref{fig:skyrm_uncomp} (for example, the second figure in the second row). A further increase of the exchange interaction leads to a collapse of parts of the skyrmions (Fig.\ref{fig:skyrm_uncomp}). A continued increase of the exchange interaction, $J^{\mathrm{FM/AFM}}$, causes the effective field to be sufficiently strong to overcome the DMI and the magnetization of FM layer will become uniform. 

Another important aspect to consider, while studying skyrmions is the impact of the DMI on the magnetic structure of the FM layer. We should note that the skyrmion lattice presented in Fig.\ref{fig:skyrm_uncomp} is very sensitive to the DMI value. As can be seen from Fig.\ref{fig:skyrm_uncomp}, the increase of DMI value in FM leads to skyrmions observation for larger values of exchange interactions through the interface, since a stronger effective field is required to overcome DMI.

\begin{figure}[h!]
\centering
\includegraphics[width=0.45\textwidth]{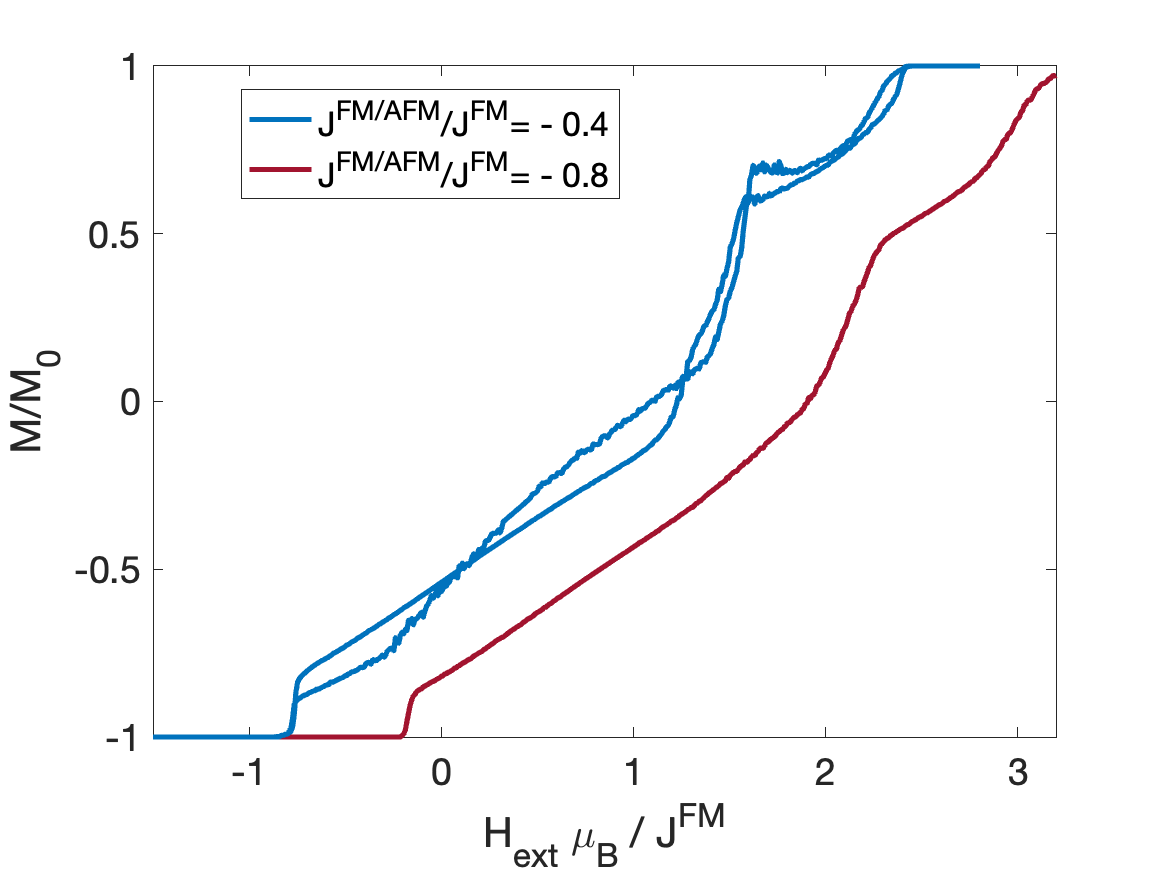}
\vspace{-0.2cm}
\caption{Exchange-biased magnetic hysteresis of a FM film for $J^{\mathrm{FM/AFM}}/J^{\mathrm{FM}}=-0.4$ (blue), and $J^{\mathrm{FM/AFM}}/J^{\mathrm{FM}}=-0.8$ (red).}
\vspace{-0.2cm}
\label{fig:hyst_uncomp}
\end{figure}

In Fig.\ref{fig:hyst_uncomp} the magnetic hysteresis loops are shown for two values of interplane exchange interactions, $J^{\mathrm{FM/AFM}}/J^{FM}$= --0.4 and --0.8. 
It can be seen that in both cases the hysteresis loops are shifted and the shift is increasing with the increase of the exchange interaction $J^{\mathrm{FM/AFM}}$. We notice that hysteresis loops consist of several loops, something that was also observed experimentally in Ref. \cite{chen2023tailoring}. At zero fields the inclined part of the hysteresis loop corresponds to non-collinear magnetization structure in the film. 

\begin{figure}[h!]
\begin{tabular}{c}
  \includegraphics[width=0.45\textwidth]{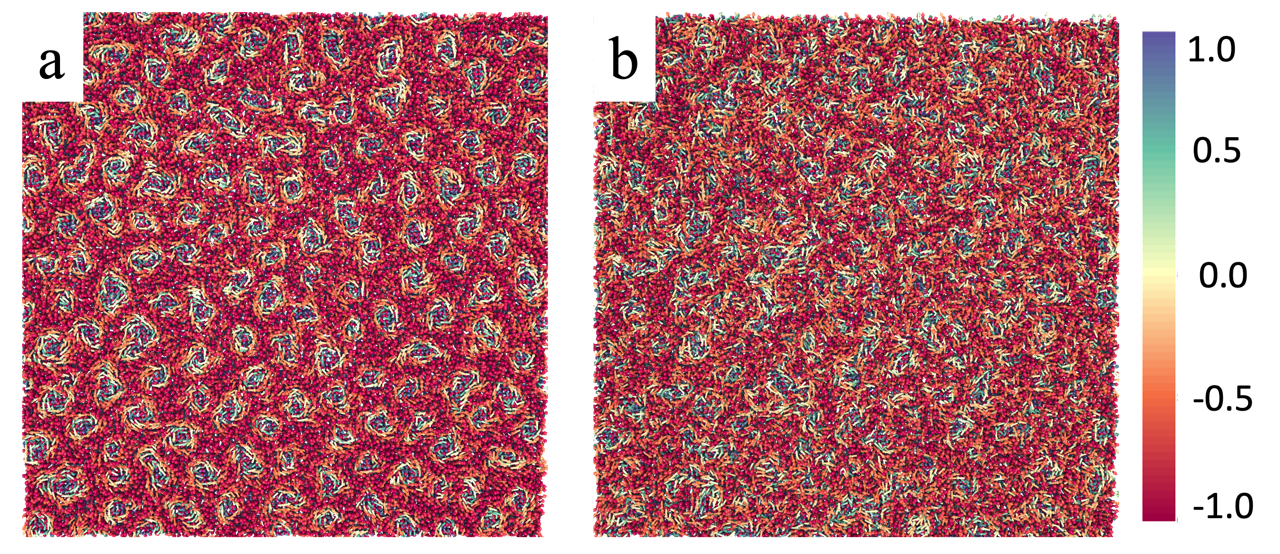}   \\
\includegraphics[width=0.40\textwidth]{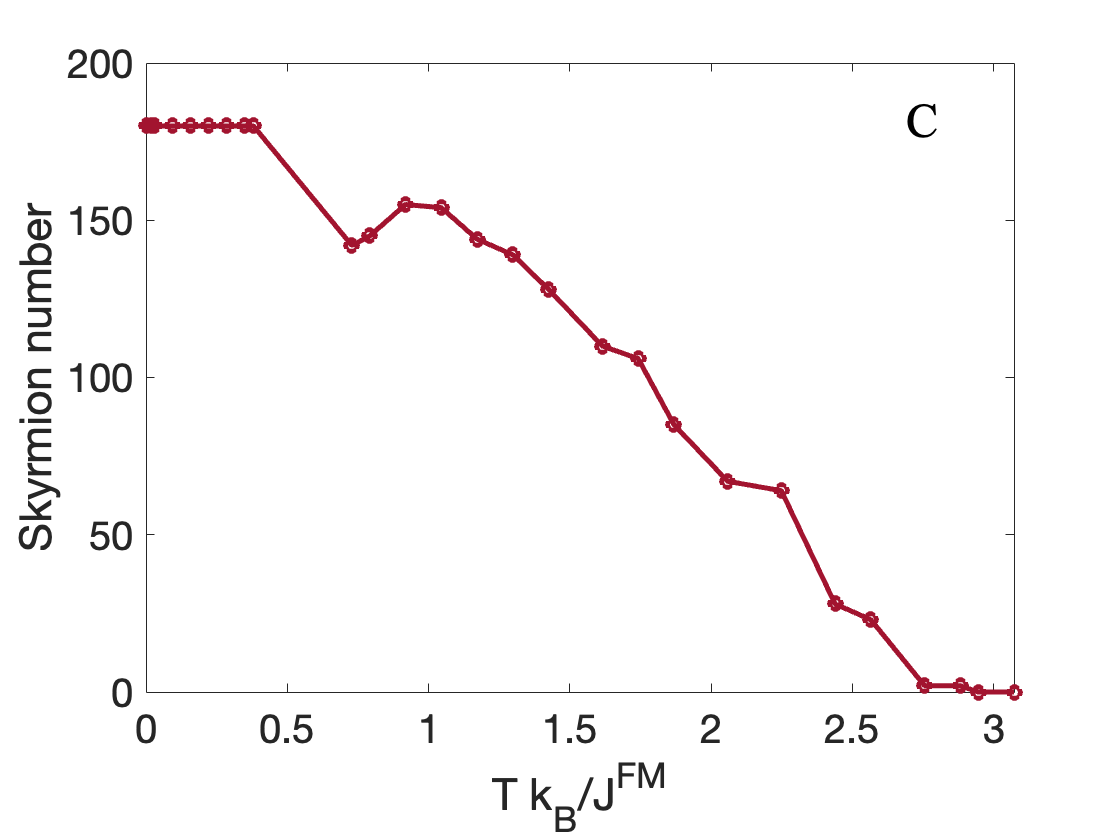}
\end{tabular}
\caption{Ferromagnetic skyrmions in FM-AFM bilayer in zero external magnetic field for different temperatures (a) $T k_B /J^{FM}=0.8$, (b)  $T k_B /J^{FM}=1.4$. Skyrmion number \cite{PhysRevB.39.7212} versus temperature (c). The value of $J^{\mathrm{FM/AFM}}/J^{\mathrm{FM}}=-0.4$.}
\vspace{-0.2cm}
\label{fig:skyrm_temp_uncomp}
\end{figure}

The results in Fig.\ref{fig:skyrm_uncomp} are obtained for very low temperatures. However, it is known that the stability and inner structure of skyrmions depend on the temperature, which is especially pronounced near the Curie temperature (please see Ref.\cite{wang2022exponential} and reference therein).  Therefore, we have studied the impact of temperature on the skyrmion lattice. For low temperatures the magnetic structure remains similar to the one presented in Fig.\ref{fig:skyrm_uncomp}, however, with the rise of the temperature the magnetic structure changes (see, for example, Fig.\ref{fig:skyrm_temp_uncomp}a) and skyrmions disappear for sufficiently high temperatures. The explicit temperature dependence of the skyrmion number of the simulation cell is presented in Fig.\ref{fig:skyrm_temp_uncomp}c, showing that the topological magnetic order disappears at a temperature just above 2.5 k$_B$/J$^{FM}$.

\subsection{Compensated FM-AFM interface}

\begin{figure}[h!]
\centering
\includegraphics[width=0.4\textwidth]{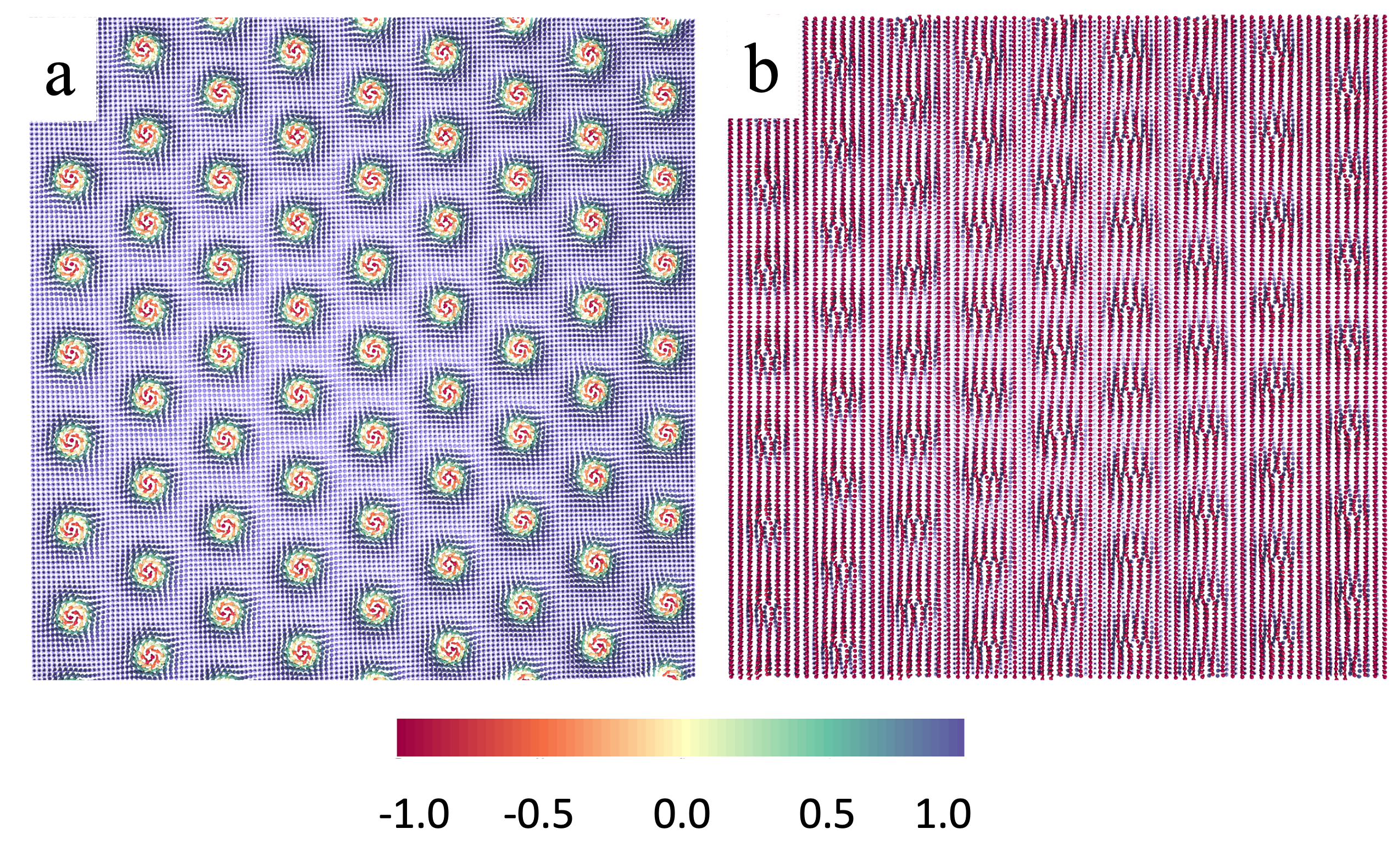}
\vspace{-0.2cm}
\caption{Magnetic structure of FM layer (a) and first (b) AFM layer correspondingly for $K^{\mathrm{AFM}}/J^{\mathrm{FM}}=-0.25$ and $|J^{\mathrm{FM/AFM}}/J^{\mathrm{FM}}|=0.6$.}
\vspace{-0.2cm}
\label{fig:skyrm_comp}
\end{figure}

Next we study the case of afully compensated AFM interface. The structure of the model can be seen in Fig.\ref{fig:model_uncom_com}(b). In this case, the total magnetic moment of the AFM interface atoms in contact with the FM layer is zero  and therefore there is no effective exchange field acting on the magnetic moments in the FM layer that stems from atoms in the AMF layer, at least if the AFM moments are frozen in perfect AFM order with vanishing fluctuations. This is expected to not result in the formation of skyrmions. However, fluctuations of magnetic moments will generate an effective exchange field across the AFM-FM interface, with the possibility to induce skyrmions. To allow for this we have for the simulations presented in this section used a magnetic anisotropy in the AFM layers that is half of that used in the uncompensated model. We present the magnetic structure of FM and AFM layers in zero magnetic field, for and lower magnetic anisotropy in the AFM layer, see Fig.\ref{fig:skyrm_comp}. In the absence of exchange through the FM-AFM interface, the magnetic structure of FM film is naturally similar to uncompensated model (see Fig.\ref{fig:skyrm_uncomp} for $J^{\mathrm{FM/AFM}}/J^{\mathrm{FM}}=0$). It can be seen from Fig.\ref{fig:skyrm_comp}(a), that there are skyrmions in the FM layer. We would like to note that no skyrmions would be observed for high values of the magnetic anisotropy, for the same values of other parameters. Moreover, if we study closely the magnetic structure of the AFM layers, as displayed in Fig.\ref{fig:skyrm_comp}(b), an imprint of FM magnetic structure, in this case, of skyrmions in the AFM layer can be observed. The transfer of the magnetic structure of FM to AFM in exchange-biased bilayers was shown experimentally in Ref.\cite{rana2021imprint}. Such random topological magnetic textures could be useful for e.g. reservoir computing \cite{PhysRevApplied.14.054020}. Finally, it was shown experimentally in \cite{rana2021imprint}, that induction of skyrmions in AFM by transferring FM magnetic order to AFM by exchange bias, is a very promising technique. Our findings support these experimental observations. Antiferromagnetic skyrmions are very promising candidates for memory devices because they are expected to move without a skyrmion Hall effect.

\subsection{Partly uncompensated interface. Impact of interface roughness}

 In this section, we study the case of a partly uncompensated interface, where uncompensated spins appear due to, for example, interface roughness or disorder. It was shown in Ref.\cite{PhysRevApplied.14.054020} that skyrmions pinned by inhomogeneities can be useful for the implementation of reservoir computing. The scheme of the model is presented in Fig.\ref{fig:model_uncom_com}(c). We start with a perfectly compensated AFM interface (checkerboard AFM pattern) and then assume that one out of four AFM magnetic moments on the interface is uncompensated (the orange one in Fig.\ref{fig:model_uncom_com}(c)). We study cases when coupling ferromagnetic $J^{\mathrm{FM/AFM}}_{\mathrm{uncomp}}>0$, or antiferromagnetic $J^{\mathrm{FM/AFM}}_{\mathrm{uncomp}}<0$ to the FM layer and to its neighbors in the layer. The other AFM layers are ordered as perfect AFM. Therefore, in this system, the EBP, and, therefore, the effective field will depend on the ratio between coupling in an AFM interface layer, and through the FM-AFM interface. 
 
\begin{figure}[h!]
\centering
\includegraphics[width=0.47\textwidth]{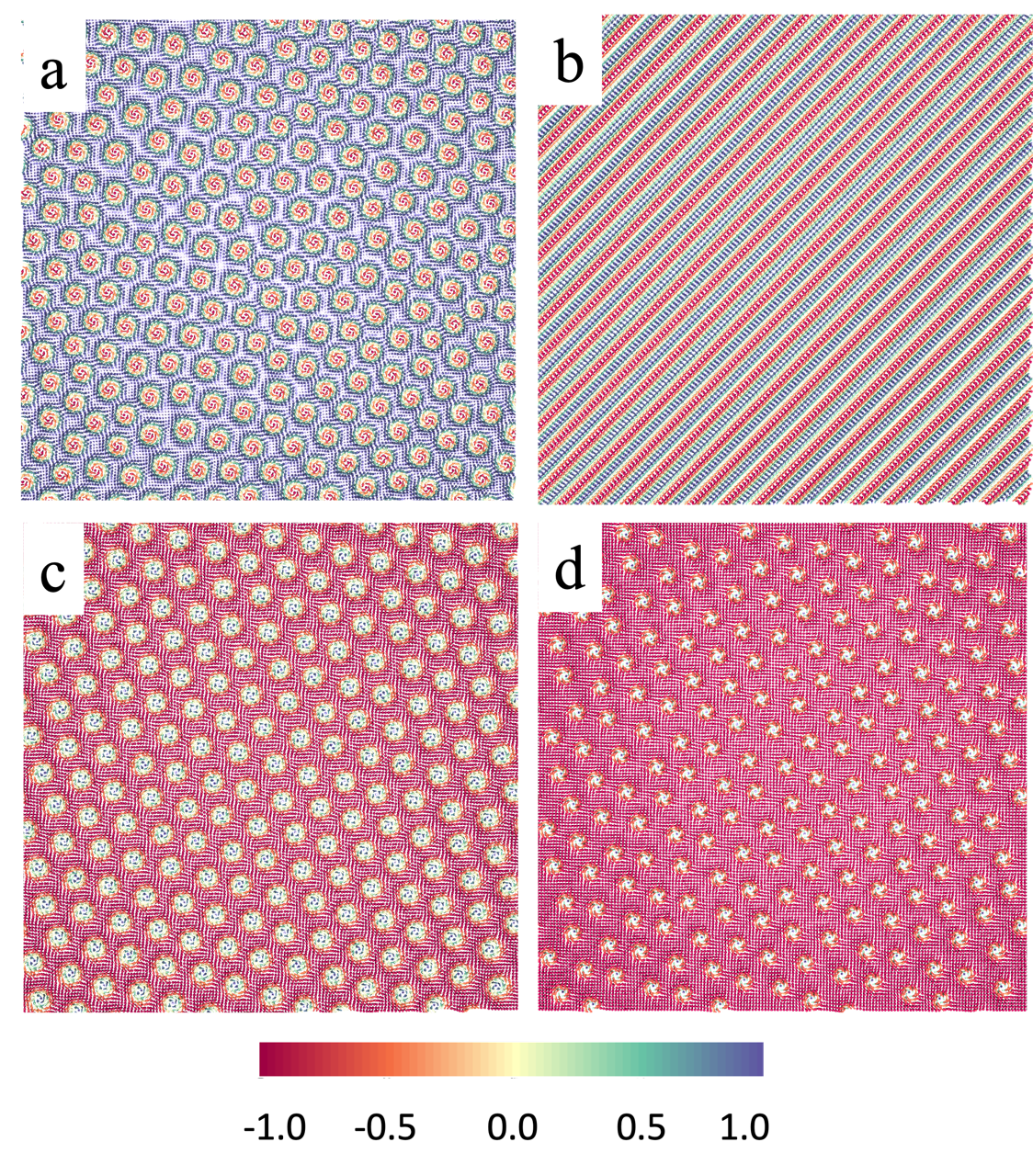}
\vspace{-0.2cm}
\caption{Ferromagnetic skyrmions in FM-AFM bilayer with partly uncompensated interface in zero external magnetic field for various values of exchange interaction $J^{\mathrm{FM/AFM}}$ through the FM-AFM interface. In particular, (a) $J^{\mathrm{FM/AFM}}_{\mathrm{uncomp}}/J^{\mathrm{FM/AFM}}_{\mathrm{comp}}=-10.5$, (b) $J^{\mathrm{FM/AFM}}_{\mathrm{uncomp}}/J^{\mathrm{FM/AFM}}_{\mathrm{comp}}=0.0$, (c) $J^{\mathrm{FM/AFM}}_{\mathrm{uncomp}}/J^{\mathrm{FM/AFM}}_{\mathrm{comp}}=30.5$, (d) $J^{\mathrm{FM/AFM}}_{\mathrm{uncomp}}/J^{\mathrm{FM/AFM}}_{\mathrm{comp}}=100.0$.}
\label{fig:skyrm_partly}
\end{figure}

 In particular, in Fig.\ref{fig:skyrm_partly} we present the case when the coupling of uncompensated AFM magnetic moments (orange ones in Fig.\ref{fig:model_uncom_com}c) through the interface is much stronger than the coupling of compensated AFM spins (green ones in Fig.\ref{fig:model_uncom_com}c) $J^{\mathrm{FM/AFM}}_{\mathrm{uncomp}}>J^{\mathrm{FM/AFM}}_{\mathrm{comp}}$. At zero exchange through the interface, similarly, to compensated or uncompensated model, we observe stripe domains in FM layer (see Fig.\ref{fig:skyrm_partly}(b)). Then, with an increase of $J^{\mathrm{FM/AFM}}_{\mathrm{uncomp}}$ exchange coupling we can observe the formation of a skyrmion lattice in the FM layer (please see Fig.\ref{fig:skyrm_partly}(c)). Skyrmions become smaller with increasing exchange interaction Fig.\ref{fig:skyrm_partly}(d) and then, skyrmions lattice collapse.  

\begin{figure}[h!]
\centering
\includegraphics[width=0.45\textwidth]{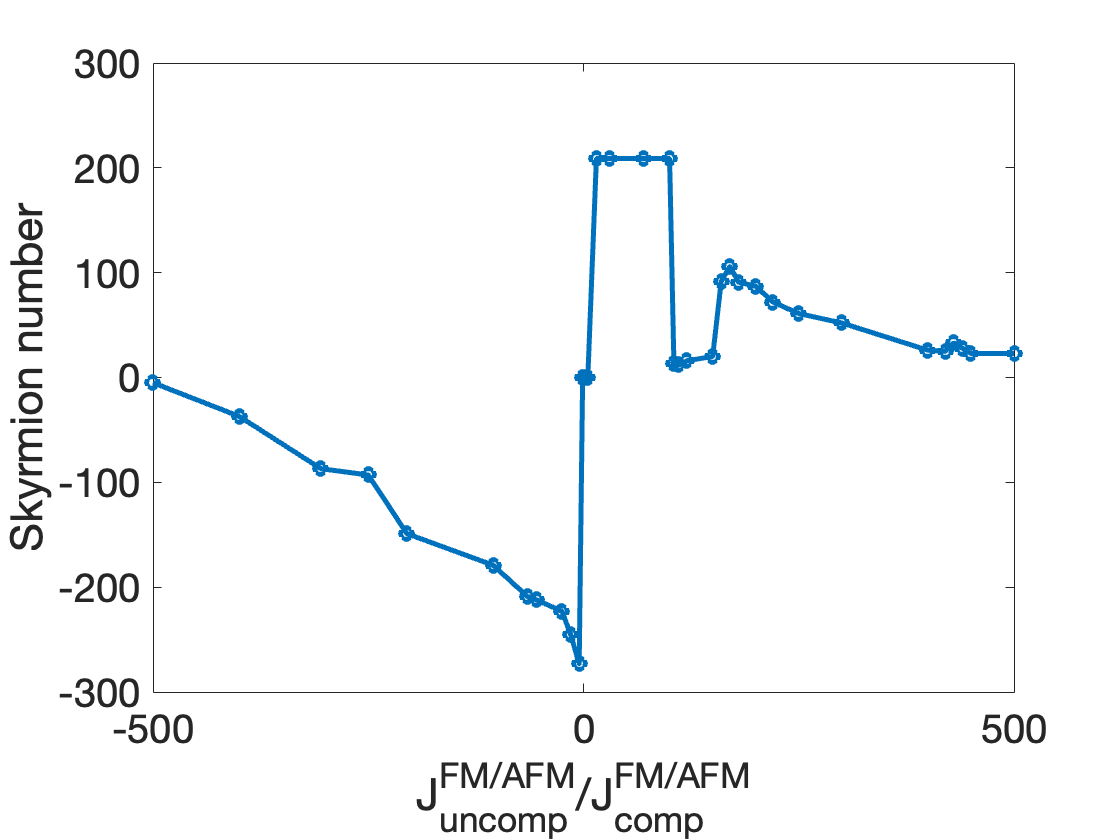}
\vspace{-0.2cm}
\caption{Skyrmion number depends on the exchange interaction ratio between compensated and uncompensated spins in the AFM interface. The minus/plus sign in the skyrmion number denotes skyrmion chirality.}
\vspace{-0.2cm}
\label{fig:skyrm_partly_numb}
\end{figure}

Above, we have studied the case of positively coupled through the FM-AFM interface uncompensated AFM spins $J^{\mathrm{FM/AFM}}_{\mathrm{uncomp}}>0$, while were $J^{\mathrm{FM/AFM}}_{\mathrm{comp}}<0$. However, one can observe skyrmions in this model even for negative values of exchange interaction through the interface of  uncompensated AFM spins $J^{\mathrm{FM/AFM}}_{\mathrm{uncomp}}<0$. For small negative values values of $J^{\mathrm{FM/AFM}}_{\mathrm{uncomp}}$ the magnetization of FM layer is similar to Fig.\ref{fig:skyrm_partly}, however, with the rise of exchange interaction value, we can observe skyrmion lattice, as presented in Fig.\ref{fig:skyrm_partly}(a). Importantly, for negative values of coupling $J^{\mathrm{FM/AFM}}_{\mathrm{uncomp}}$  skyrmions change chirality, and one can observe a lattice of skyrmions (Fig.\ref{fig:skyrm_partly}(a)), with negative chirality. With further increase of coupling number of skyrmions decrease (Fig.\ref{fig:skyrm_partly}b) and then they disappear. The dependence of skyrmions number as a function of exchange interaction $J^{\mathrm{FM/AFM}}_{\mathrm{uncomp}}$ is presented in Fig.\ref{fig:skyrm_partly_numb}. One can observe, that there is the critical value of exchange interaction allowing skyrmions formation, however, skyrmions are observed in a wide range of exchange interactions through the interface, both positive and negative ones. We can conclude, that with only 25$\%$ of spins uncompensated at the AFM interface, it takes enormous couplings to saturate magnetization in an FM film, and therefore, skyrmions exist in a much wider range of exchange interaction values than in the case of fully uncompensated interface considered above. 

\subsection{Random interface. Impact of interface roughness}

In the previous section, we studied a model when a perfectly compensated interface became partly uncompensated due to, for example, disorder. In that case, the uncompensated spins were distributed periodically at the interface. However, in real systems with rough/disordered interfaces, these defects would be distributed randomly.

In this section, we demonstrate the impact of interface roughness/disorder by introducing randomly distributed magnetic moments with positive and negative exchange interaction through the interface. The case when all of the spins have $J^{\mathrm{FM/AFM}} >0$ (or $J^{\mathrm{FM/AFM}} <0$) would correspond to the uncompensated interface, studied above.  In turn, the situation when the number of spins with $J^{\mathrm{FM/AFM}} >0$ and $J^{\mathrm{FM/AFM}} <0$ is the same corresponds to a compensated interface. However, unlike for the perfectly compensated interface above, here, spins from two AFM sublattices are distributed randomly, and therefore, effective fields can arise and lead to skyrmions stabilization. The other ratio between two AFM sublattices will lead to cases close to the partly uncompensated interface. 

\begin{figure}[htb]
\centering
\includegraphics[width=0.45\textwidth]{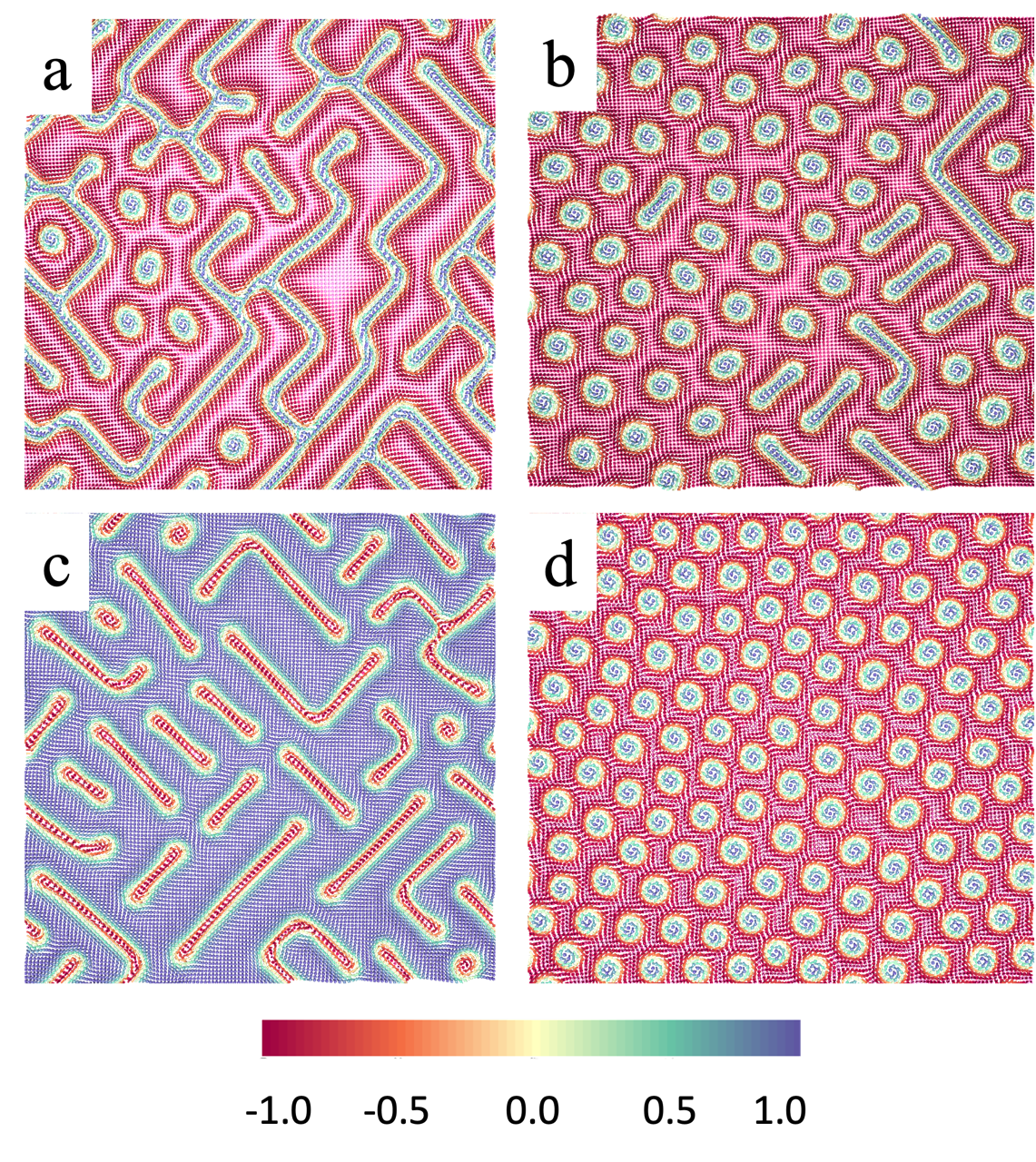}
\vspace{-0.2cm}
\caption{Ferromagnetic skyrmions in FM-AFM bilayer for rough interface with various percentage of spins from two AFM sublattices. In particular, (a) 10$\%$ of $J^{\mathrm{FM/AFM}}>0$, 90$\%$  of $J^{\mathrm{FM/AFM}}<0$ (b) 50$\%$ of $J^{\mathrm{FM/AFM}}>0$, 50$\%$  of$J^{\mathrm{FM/AFM}}<0$ (c) 80$\%$ of $J^{\mathrm{FM/AFM}}>0$, 20$\%$ of $J^{\mathrm{FM/AFM}}<0$ (d) 90$\%$ of $J^{\mathrm{FM/AFM}}>0$,  10$\%$ of $J^{\mathrm{FM/AFM}}<0$. The values of $|J^{\mathrm{FM/AFM}}/J^{\mathrm{FM}}|=0.4$ for both AFM sublattices. }
\vspace{-0.2cm}
\label{fig:skyrm_rough}
\end{figure}

The magnetic structure of a FM layer is presented in Fig.\ref{fig:skyrm_rough}. We study various percentage of spins from two AFM sublattices presented at the AFM interface layer, which corresponds to various percentage of positively of negatively coupled spins at the interface. In Fig.\ref{fig:skyrm_rough}a,c,d we illustrate the cases when most of the spins either negative or positively coupled. It can be seen from the figure that the behavior of the system is not symmetric, concerning the amount of positive (negative) coupling through the AFM-AFM interface. In particular, one can observe the coexistence of labyrinth domains and skyrmions for case Fig.\ref{fig:skyrm_rough}(a), however, in case Fig.\ref{fig:skyrm_rough}(d) we obtain stable skyrmions lattice. It is due to the sign of DMI, and also to the exchange bias effect itself.  In Fig.\ref{fig:skyrm_rough}(b) we present the stabilization of skyrmions in FM-AFM bilayer, when the amount of positively and negatively coupled spins at the interface is the same. In all presented cases, one can observe the imprint of FM magnetic structure in the AFM interface layer, for example, skyrmions lattice from Fig.\ref{fig:skyrm_rough}(d) will be imprinted in the AFM interface layer, leading to the similar periodic pattern, with a bigger deviation of AFM magnetic moments in places corresponding to skyrmions positions. 

\section{Conclusions}

We have performed atomistic spin dynamic simulations to study the stabilization of magnetic skyrmions in an exchange-biased FM-AFM bilayer in a zero external magnetic field. We have investigated the cases of compensated, uncompensated FM-AFM interfaces, and the impact of geometrical roughness at the interface. We have shown that, in the case of an uncompensated interface, the effective field acting from AFM to FM film leads to the appearance of a skyrmions and even a skyrmion lattice. However, either too weak or too strong an exchange through the FM-AFM interface destroys skyrmions. The fully compensated AFM interface results in the appearance of skyrmions, however, only for low values of magnetic anisotropy in AFM. Moreover, due to the transfer by exchange bias of the magnetic structure of a FM to an AFM, one can observe imprint of FM skyrmions in AFM, which was also observed experimentally. In the case of a partly uncompensated interface, where only 25$\%$ of the spins at the AFM interface are uncompensated, due to competition between compensated and uncompensated spins, skyrmions can be observed both for positive and negative exchange interaction through the FM-AFM interface in the wide range of exchange values and skyrmions chirality depend of the exchange interactions through the FM-AFM interface. We show that the interface roughness can lead to skyrmions lattice stabilization in the FM layer for the cases of both compensated and uncompensated interfaces.
The results of this work could be useful in reducing the energy consumption of perspective memory devices based on skyrmions and for skyrmion-based reservoir computing applications.

\section{Data availability}

The data used to produce results presented in paper, are obtained using UppASD software, available to download, from the UppASD web-page \cite{UppASD}.

\section{ACKNOWLEDGEMENTS}

This work was financially supported by Olle Engkvist foundation and Knut och Alice Wallenbergs Stiftelse, contract 2018.0060. Computations were enabled by resources provided by the Swedish National Infrastructure for Computing (SNIC) at NSC, partially funded by the Swedish Research Council. 
O.E. acknowledge support from 
the Wallenberg Initiative Materials Science
for Sustainability (WISE) funded by the Knut and Alice
Wallenberg Foundation (KAW), the Swedish Research Council (VR), and STandUPP. A.B. and O.E. acknowledge support from eSSENCE. M.P. acknowledge support from WISE-WASP initiative.

\nocite{*}

\bibliography{apssamp}

\providecommand{\noopsort}[1]{}\providecommand{\singleletter}[1]{#1}%
\begin{thebibliography}{34}%
\makeatletter
\providecommand \@ifxundefined [1]{%
 \@ifx{#1\undefined}
}%
\providecommand \@ifnum [1]{%
 \ifnum #1\expandafter \@firstoftwo
 \else \expandafter \@secondoftwo
 \fi
}%
\providecommand \@ifx [1]{%
 \ifx #1\expandafter \@firstoftwo
 \else \expandafter \@secondoftwo
 \fi
}%
\providecommand \natexlab [1]{#1}%
\providecommand \enquote  [1]{``#1''}%
\providecommand \bibnamefont  [1]{#1}%
\providecommand \bibfnamefont [1]{#1}%
\providecommand \citenamefont [1]{#1}%
\providecommand \href@noop [0]{\@secondoftwo}%
\providecommand \href [0]{\begingroup \@sanitize@url \@href}%
\providecommand \@href[1]{\@@startlink{#1}\@@href}%
\providecommand \@@href[1]{\endgroup#1\@@endlink}%
\providecommand \@sanitize@url [0]{\catcode `\\12\catcode `\$12\catcode
  `\&12\catcode `\#12\catcode `\^12\catcode `\_12\catcode `\%12\relax}%
\providecommand \@@startlink[1]{}%
\providecommand \@@endlink[0]{}%
\providecommand \url  [0]{\begingroup\@sanitize@url \@url }%
\providecommand \@url [1]{\endgroup\@href {#1}{\urlprefix }}%
\providecommand \urlprefix  [0]{URL }%
\providecommand \Eprint [0]{\href }%
\providecommand \doibase [0]{https://doi.org/}%
\providecommand \selectlanguage [0]{\@gobble}%
\providecommand \bibinfo  [0]{\@secondoftwo}%
\providecommand \bibfield  [0]{\@secondoftwo}%
\providecommand \translation [1]{[#1]}%
\providecommand \BibitemOpen [0]{}%
\providecommand \bibitemStop [0]{}%
\providecommand \bibitemNoStop [0]{.\EOS\space}%
\providecommand \EOS [0]{\spacefactor3000\relax}%
\providecommand \BibitemShut  [1]{\csname bibitem#1\endcsname}%
\let\auto@bib@innerbib\@empty
\bibitem [{\citenamefont {Tokura}\ and\ \citenamefont
  {Kanazawa}(2020)}]{tokura2020magnetic}%
  \BibitemOpen
  \bibfield  {author} {\bibinfo {author} {\bibfnamefont {Y.}~\bibnamefont
  {Tokura}}\ and\ \bibinfo {author} {\bibfnamefont {N.}~\bibnamefont
  {Kanazawa}},\ }\bibfield  {title} {\bibinfo {title} {Magnetic skyrmion
  materials},\ }\href@noop {} {\bibfield  {journal} {\bibinfo  {journal}
  {Chemical Reviews}\ }\textbf {\bibinfo {volume} {121}},\ \bibinfo {pages}
  {2857} (\bibinfo {year} {2020})}\BibitemShut {NoStop}%
\bibitem [{\citenamefont {Wang}\ \emph
  {et~al.}(2022{\natexlab{a}})\citenamefont {Wang}, \citenamefont
  {Bheemarasetty}, \citenamefont {Duan}, \citenamefont {Zhou},\ and\
  \citenamefont {Xiao}}]{WANG2022169905}%
  \BibitemOpen
  \bibfield  {author} {\bibinfo {author} {\bibfnamefont {K.}~\bibnamefont
  {Wang}}, \bibinfo {author} {\bibfnamefont {V.}~\bibnamefont {Bheemarasetty}},
  \bibinfo {author} {\bibfnamefont {J.}~\bibnamefont {Duan}}, \bibinfo {author}
  {\bibfnamefont {S.}~\bibnamefont {Zhou}},\ and\ \bibinfo {author}
  {\bibfnamefont {G.}~\bibnamefont {Xiao}},\ }\bibfield  {title} {\bibinfo
  {title} {Fundamental physics and applications of skyrmions: A review},\
  }\href {https://doi.org/https://doi.org/10.1016/j.jmmm.2022.169905}
  {\bibfield  {journal} {\bibinfo  {journal} {Journal of Magnetism and Magnetic
  Materials}\ }\textbf {\bibinfo {volume} {563}},\ \bibinfo {pages} {169905}
  (\bibinfo {year} {2022}{\natexlab{a}})}\BibitemShut {NoStop}%
\bibitem [{\citenamefont {Everschor-Sitte}\ \emph {et~al.}(2018)\citenamefont
  {Everschor-Sitte}, \citenamefont {Masell}, \citenamefont {Reeve},\ and\
  \citenamefont {Kl{\"a}ui}}]{everschor2018perspective}%
  \BibitemOpen
  \bibfield  {author} {\bibinfo {author} {\bibfnamefont {K.}~\bibnamefont
  {Everschor-Sitte}}, \bibinfo {author} {\bibfnamefont {J.}~\bibnamefont
  {Masell}}, \bibinfo {author} {\bibfnamefont {R.~M.}\ \bibnamefont {Reeve}},\
  and\ \bibinfo {author} {\bibfnamefont {M.}~\bibnamefont {Kl{\"a}ui}},\
  }\bibfield  {title} {\bibinfo {title} {Perspective: Magnetic
  skyrmions—overview of recent progress in an active research field},\
  }\href@noop {} {\bibfield  {journal} {\bibinfo  {journal} {Journal of Applied
  Physics}\ }\textbf {\bibinfo {volume} {124}} (\bibinfo {year}
  {2018})}\BibitemShut {NoStop}%
\bibitem [{\citenamefont {Ajejas}\ \emph {et~al.}(2023)\citenamefont {Ajejas},
  \citenamefont {Sassi}, \citenamefont {Legrand}, \citenamefont {Srivastava},
  \citenamefont {Collin}, \citenamefont {Vecchiola}, \citenamefont
  {Bouzehouane}, \citenamefont {Reyren},\ and\ \citenamefont
  {Cros}}]{ajejas2023densely}%
  \BibitemOpen
  \bibfield  {author} {\bibinfo {author} {\bibfnamefont {F.}~\bibnamefont
  {Ajejas}}, \bibinfo {author} {\bibfnamefont {Y.}~\bibnamefont {Sassi}},
  \bibinfo {author} {\bibfnamefont {W.}~\bibnamefont {Legrand}}, \bibinfo
  {author} {\bibfnamefont {T.}~\bibnamefont {Srivastava}}, \bibinfo {author}
  {\bibfnamefont {S.}~\bibnamefont {Collin}}, \bibinfo {author} {\bibfnamefont
  {A.}~\bibnamefont {Vecchiola}}, \bibinfo {author} {\bibfnamefont
  {K.}~\bibnamefont {Bouzehouane}}, \bibinfo {author} {\bibfnamefont
  {N.}~\bibnamefont {Reyren}},\ and\ \bibinfo {author} {\bibfnamefont
  {V.}~\bibnamefont {Cros}},\ }\bibfield  {title} {\bibinfo {title} {Densely
  packed skyrmions stabilized at zero magnetic field by indirect exchange
  coupling in multilayers},\ }\href@noop {} {\bibfield  {journal} {\bibinfo
  {journal} {APL Materials}\ }\textbf {\bibinfo {volume} {11}} (\bibinfo {year}
  {2023})}\BibitemShut {NoStop}%
\bibitem [{\citenamefont {Jiang}\ \emph {et~al.}(2017)\citenamefont {Jiang},
  \citenamefont {Chen}, \citenamefont {Liu}, \citenamefont {Zang},
  \citenamefont {{te Velthuis}},\ and\ \citenamefont {Hoffmann}}]{JIANG20171}%
  \BibitemOpen
  \bibfield  {author} {\bibinfo {author} {\bibfnamefont {W.}~\bibnamefont
  {Jiang}}, \bibinfo {author} {\bibfnamefont {G.}~\bibnamefont {Chen}},
  \bibinfo {author} {\bibfnamefont {K.}~\bibnamefont {Liu}}, \bibinfo {author}
  {\bibfnamefont {J.}~\bibnamefont {Zang}}, \bibinfo {author} {\bibfnamefont
  {S.~G.}\ \bibnamefont {{te Velthuis}}},\ and\ \bibinfo {author}
  {\bibfnamefont {A.}~\bibnamefont {Hoffmann}},\ }\bibfield  {title} {\bibinfo
  {title} {Skyrmions in magnetic multilayers},\ }\href
  {https://doi.org/https://doi.org/10.1016/j.physrep.2017.08.001} {\bibfield
  {journal} {\bibinfo  {journal} {Physics Reports}\ }\textbf {\bibinfo {volume}
  {704}},\ \bibinfo {pages} {1} (\bibinfo {year} {2017})},\ \bibinfo {note}
  {skyrmions in Magnetic Multilayers}\BibitemShut {NoStop}%
\bibitem [{\citenamefont {Chen}\ \emph {et~al.}(2023)\citenamefont {Chen},
  \citenamefont {Tai}, \citenamefont {Tan}, \citenamefont {Tan}, \citenamefont
  {Lim}, \citenamefont {Ho},\ and\ \citenamefont
  {Soumyanarayanan}}]{chen2023tailoring}%
  \BibitemOpen
  \bibfield  {author} {\bibinfo {author} {\bibfnamefont {X.}~\bibnamefont
  {Chen}}, \bibinfo {author} {\bibfnamefont {T.}~\bibnamefont {Tai}}, \bibinfo
  {author} {\bibfnamefont {H.~R.}\ \bibnamefont {Tan}}, \bibinfo {author}
  {\bibfnamefont {H.~K.}\ \bibnamefont {Tan}}, \bibinfo {author} {\bibfnamefont
  {R.}~\bibnamefont {Lim}}, \bibinfo {author} {\bibfnamefont {P.}~\bibnamefont
  {Ho}},\ and\ \bibinfo {author} {\bibfnamefont {A.}~\bibnamefont
  {Soumyanarayanan}},\ }\bibfield  {title} {\bibinfo {title} {Tailoring
  zero-field magnetic skyrmions in chiral multilayers by a duet of interlayer
  exchange couplings},\ }\href@noop {} {\bibfield  {journal} {\bibinfo
  {journal} {arXiv preprint arXiv:2301.07327}\ } (\bibinfo {year}
  {2023})}\BibitemShut {NoStop}%
\bibitem [{\citenamefont {Rana}\ \emph {et~al.}(2021)\citenamefont {Rana},
  \citenamefont {Lopes~Seeger}, \citenamefont {Ruiz-G{\'o}mez}, \citenamefont
  {Juge}, \citenamefont {Zhang}, \citenamefont {Bairagi}, \citenamefont {Pham},
  \citenamefont {Belmeguenai}, \citenamefont {Auffret}, \citenamefont
  {Foerster} \emph {et~al.}}]{rana2021imprint}%
  \BibitemOpen
  \bibfield  {author} {\bibinfo {author} {\bibfnamefont {K.~G.}\ \bibnamefont
  {Rana}}, \bibinfo {author} {\bibfnamefont {R.}~\bibnamefont {Lopes~Seeger}},
  \bibinfo {author} {\bibfnamefont {S.}~\bibnamefont {Ruiz-G{\'o}mez}},
  \bibinfo {author} {\bibfnamefont {R.}~\bibnamefont {Juge}}, \bibinfo {author}
  {\bibfnamefont {Q.}~\bibnamefont {Zhang}}, \bibinfo {author} {\bibfnamefont
  {K.}~\bibnamefont {Bairagi}}, \bibinfo {author} {\bibfnamefont {V.~T.}\
  \bibnamefont {Pham}}, \bibinfo {author} {\bibfnamefont {M.}~\bibnamefont
  {Belmeguenai}}, \bibinfo {author} {\bibfnamefont {S.}~\bibnamefont
  {Auffret}}, \bibinfo {author} {\bibfnamefont {M.}~\bibnamefont {Foerster}},
  \emph {et~al.},\ }\bibfield  {title} {\bibinfo {title} {Imprint from
  ferromagnetic skyrmions in an antiferromagnet via exchange bias},\
  }\href@noop {} {\bibfield  {journal} {\bibinfo  {journal} {Applied Physics
  Letters}\ }\textbf {\bibinfo {volume} {119}} (\bibinfo {year}
  {2021})}\BibitemShut {NoStop}%
\bibitem [{\citenamefont {Mallick}\ \emph {et~al.}(2022)\citenamefont
  {Mallick}, \citenamefont {Panigrahy}, \citenamefont {Pradhan},\ and\
  \citenamefont {Rohart}}]{mallick2022current}%
  \BibitemOpen
  \bibfield  {author} {\bibinfo {author} {\bibfnamefont {S.}~\bibnamefont
  {Mallick}}, \bibinfo {author} {\bibfnamefont {S.}~\bibnamefont {Panigrahy}},
  \bibinfo {author} {\bibfnamefont {G.}~\bibnamefont {Pradhan}},\ and\ \bibinfo
  {author} {\bibfnamefont {S.}~\bibnamefont {Rohart}},\ }\bibfield  {title}
  {\bibinfo {title} {Current-induced nucleation and motion of skyrmions in zero
  magnetic field},\ }\href@noop {} {\bibfield  {journal} {\bibinfo  {journal}
  {Physical Review Applied}\ }\textbf {\bibinfo {volume} {18}},\ \bibinfo
  {pages} {064072} (\bibinfo {year} {2022})}\BibitemShut {NoStop}%
\bibitem [{\citenamefont {Brand{\~a}o}\ \emph {et~al.}(2019)\citenamefont
  {Brand{\~a}o}, \citenamefont {Dugato}, \citenamefont {Seeger}, \citenamefont
  {Denardin}, \citenamefont {Mori},\ and\ \citenamefont
  {Cezar}}]{brandao2019observation}%
  \BibitemOpen
  \bibfield  {author} {\bibinfo {author} {\bibfnamefont {J.}~\bibnamefont
  {Brand{\~a}o}}, \bibinfo {author} {\bibfnamefont {D.}~\bibnamefont {Dugato}},
  \bibinfo {author} {\bibfnamefont {R.}~\bibnamefont {Seeger}}, \bibinfo
  {author} {\bibfnamefont {J.}~\bibnamefont {Denardin}}, \bibinfo {author}
  {\bibfnamefont {T.}~\bibnamefont {Mori}},\ and\ \bibinfo {author}
  {\bibfnamefont {J.}~\bibnamefont {Cezar}},\ }\bibfield  {title} {\bibinfo
  {title} {Observation of magnetic skyrmions in unpatterned symmetric
  multilayers at room temperature and zero magnetic field},\ }\href@noop {}
  {\bibfield  {journal} {\bibinfo  {journal} {Scientific reports}\ }\textbf
  {\bibinfo {volume} {9}},\ \bibinfo {pages} {4144} (\bibinfo {year}
  {2019})}\BibitemShut {NoStop}%
\bibitem [{\citenamefont {Mohylna}\ \emph {et~al.}(2022)\citenamefont
  {Mohylna}, \citenamefont {Tkachenko},\ and\ \citenamefont
  {Žukovič}}]{MOHYLNA2022128350}%
  \BibitemOpen
  \bibfield  {author} {\bibinfo {author} {\bibfnamefont {M.}~\bibnamefont
  {Mohylna}}, \bibinfo {author} {\bibfnamefont {V.}~\bibnamefont {Tkachenko}},\
  and\ \bibinfo {author} {\bibfnamefont {M.}~\bibnamefont {Žukovič}},\
  }\bibfield  {title} {\bibinfo {title} {Road to zero-field antiferromagnetic
  skyrmions in a frustrated afm/fm heterostructure},\ }\href
  {https://doi.org/https://doi.org/10.1016/j.physleta.2022.128350} {\bibfield
  {journal} {\bibinfo  {journal} {Physics Letters A}\ }\textbf {\bibinfo
  {volume} {449}},\ \bibinfo {pages} {128350} (\bibinfo {year}
  {2022})}\BibitemShut {NoStop}%
\bibitem [{\citenamefont {Yu}\ \emph {et~al.}(2018)\citenamefont {Yu},
  \citenamefont {Jenkins}, \citenamefont {Ma}, \citenamefont {Razavi},
  \citenamefont {He}, \citenamefont {Yin}, \citenamefont {Shao}, \citenamefont
  {He}, \citenamefont {Wu}, \citenamefont {Li} \emph {et~al.}}]{yu2018room}%
  \BibitemOpen
  \bibfield  {author} {\bibinfo {author} {\bibfnamefont {G.}~\bibnamefont
  {Yu}}, \bibinfo {author} {\bibfnamefont {A.}~\bibnamefont {Jenkins}},
  \bibinfo {author} {\bibfnamefont {X.}~\bibnamefont {Ma}}, \bibinfo {author}
  {\bibfnamefont {S.~A.}\ \bibnamefont {Razavi}}, \bibinfo {author}
  {\bibfnamefont {C.}~\bibnamefont {He}}, \bibinfo {author} {\bibfnamefont
  {G.}~\bibnamefont {Yin}}, \bibinfo {author} {\bibfnamefont {Q.}~\bibnamefont
  {Shao}}, \bibinfo {author} {\bibfnamefont {Q.~L.}\ \bibnamefont {He}},
  \bibinfo {author} {\bibfnamefont {H.}~\bibnamefont {Wu}}, \bibinfo {author}
  {\bibfnamefont {W.}~\bibnamefont {Li}}, \emph {et~al.},\ }\href@noop {}
  {\bibinfo {title} {Room-temperature skyrmions in an antiferromagnet-based
  heterostructure}} (\bibinfo {year} {2018})\BibitemShut {NoStop}%
\bibitem [{\citenamefont {Rana}\ \emph {et~al.}(2020)\citenamefont {Rana},
  \citenamefont {Finco}, \citenamefont {Fabre}, \citenamefont {Chouaieb},
  \citenamefont {Haykal}, \citenamefont {Buda-Prejbeanu}, \citenamefont
  {Fruchart}, \citenamefont {Le~Denmat}, \citenamefont {David}, \citenamefont
  {Belmeguenai} \emph {et~al.}}]{rana2020room}%
  \BibitemOpen
  \bibfield  {author} {\bibinfo {author} {\bibfnamefont {K.~G.}\ \bibnamefont
  {Rana}}, \bibinfo {author} {\bibfnamefont {A.}~\bibnamefont {Finco}},
  \bibinfo {author} {\bibfnamefont {F.}~\bibnamefont {Fabre}}, \bibinfo
  {author} {\bibfnamefont {S.}~\bibnamefont {Chouaieb}}, \bibinfo {author}
  {\bibfnamefont {A.}~\bibnamefont {Haykal}}, \bibinfo {author} {\bibfnamefont
  {L.~D.}\ \bibnamefont {Buda-Prejbeanu}}, \bibinfo {author} {\bibfnamefont
  {O.}~\bibnamefont {Fruchart}}, \bibinfo {author} {\bibfnamefont
  {S.}~\bibnamefont {Le~Denmat}}, \bibinfo {author} {\bibfnamefont
  {P.}~\bibnamefont {David}}, \bibinfo {author} {\bibfnamefont
  {M.}~\bibnamefont {Belmeguenai}}, \emph {et~al.},\ }\bibfield  {title}
  {\bibinfo {title} {Room-temperature skyrmions at zero field in
  exchange-biased ultrathin films},\ }\href@noop {} {\bibfield  {journal}
  {\bibinfo  {journal} {Physical review applied}\ }\textbf {\bibinfo {volume}
  {13}},\ \bibinfo {pages} {044079} (\bibinfo {year} {2020})}\BibitemShut
  {NoStop}%
\bibitem [{\citenamefont {Dieny}\ \emph {et~al.}(2024)\citenamefont {Dieny},
  \citenamefont {Fruchart},\ and\ \citenamefont
  {Marinero}}]{dieny2024comparison}%
  \BibitemOpen
  \bibfield  {author} {\bibinfo {author} {\bibfnamefont {B.}~\bibnamefont
  {Dieny}}, \bibinfo {author} {\bibfnamefont {O.}~\bibnamefont {Fruchart}},\
  and\ \bibinfo {author} {\bibfnamefont {E.~E.}\ \bibnamefont {Marinero}},\
  }\bibfield  {title} {\bibinfo {title} {Comparison of stable spin textures in
  in-plane vs. out-of-plane magnetized exchange-biased multilayers},\
  }\href@noop {} {\bibfield  {journal} {\bibinfo  {journal} {Journal of Physics
  D: Applied Physics}\ } (\bibinfo {year} {2024})}\BibitemShut {NoStop}%
\bibitem [{\citenamefont {Meiklejohn}\ and\ \citenamefont
  {Bean}(1957)}]{PhysRev.105.904}%
  \BibitemOpen
  \bibfield  {author} {\bibinfo {author} {\bibfnamefont {W.~H.}\ \bibnamefont
  {Meiklejohn}}\ and\ \bibinfo {author} {\bibfnamefont {C.~P.}\ \bibnamefont
  {Bean}},\ }\bibfield  {title} {\bibinfo {title} {New magnetic anisotropy},\
  }\href {https://doi.org/10.1103/PhysRev.105.904} {\bibfield  {journal}
  {\bibinfo  {journal} {Phys. Rev.}\ }\textbf {\bibinfo {volume} {105}},\
  \bibinfo {pages} {904} (\bibinfo {year} {1957})}\BibitemShut {NoStop}%
\bibitem [{\citenamefont {Pankratova}\ and\ \citenamefont
  {{\v{Z}}ukovi{\v{c}}}(2017)}]{pankratova2017magnetization}%
  \BibitemOpen
  \bibfield  {author} {\bibinfo {author} {\bibfnamefont {M.}~\bibnamefont
  {Pankratova}}\ and\ \bibinfo {author} {\bibfnamefont {M.}~\bibnamefont
  {{\v{Z}}ukovi{\v{c}}}},\ }\bibfield  {title} {\bibinfo {title} {Magnetization
  curves of geometrically frustrated exchange-biased fm/afm bilayers},\
  }\bibfield  {journal} {\bibinfo  {journal} {Acta Physica Polonica A}\
  }\textbf {\bibinfo {volume} {131}},\ \href
  {https://doi.org/10.12693/APhysPolA.131.642} {10.12693/APhysPolA.131.642}
  (\bibinfo {year} {2017})\BibitemShut {NoStop}%
\bibitem [{\citenamefont {Grechnev}\ \emph {et~al.}(2012)\citenamefont
  {Grechnev}, \citenamefont {Kovalev},\ and\ \citenamefont
  {Pankratova}}]{grechnev2012influence}%
  \BibitemOpen
  \bibfield  {author} {\bibinfo {author} {\bibfnamefont {A.}~\bibnamefont
  {Grechnev}}, \bibinfo {author} {\bibfnamefont {A.}~\bibnamefont {Kovalev}},\
  and\ \bibinfo {author} {\bibfnamefont {M.}~\bibnamefont {Pankratova}},\
  }\bibfield  {title} {\bibinfo {title} {Influence of magnetic anisotropy on
  hysteresis behavior in the two-spin model of a ferro/antiferromagnet bilayer
  with exchange bias},\ }\href@noop {} {\bibfield  {journal} {\bibinfo
  {journal} {Low Temperature Physics}\ }\textbf {\bibinfo {volume} {38}},\
  \bibinfo {pages} {937} (\bibinfo {year} {2012})}\BibitemShut {NoStop}%
\bibitem [{\citenamefont {Kiwi}\ \emph {et~al.}(1999)\citenamefont {Kiwi},
  \citenamefont {Mej{\'i}a-L{\'o}pez}, \citenamefont {Portugal},\ and\
  \citenamefont {Ram{\'i}rez}}]{kiwi1999exchange}%
  \BibitemOpen
  \bibfield  {author} {\bibinfo {author} {\bibfnamefont {M.}~\bibnamefont
  {Kiwi}}, \bibinfo {author} {\bibfnamefont {J.}~\bibnamefont
  {Mej{\'i}a-L{\'o}pez}}, \bibinfo {author} {\bibfnamefont {R.~D.}\
  \bibnamefont {Portugal}},\ and\ \bibinfo {author} {\bibfnamefont
  {R.}~\bibnamefont {Ram{\'i}rez}},\ }\bibfield  {title} {\bibinfo {title}
  {Exchange-bias systems with compensated interfaces},\ }\href@noop {}
  {\bibfield  {journal} {\bibinfo  {journal} {Applied Physics Letters}\
  }\textbf {\bibinfo {volume} {75}},\ \bibinfo {pages} {3995} (\bibinfo {year}
  {1999})}\BibitemShut {NoStop}%
\bibitem [{\citenamefont {Koon}(1997)}]{PhysRevLett.78.4865}%
  \BibitemOpen
  \bibfield  {author} {\bibinfo {author} {\bibfnamefont {N.~C.}\ \bibnamefont
  {Koon}},\ }\bibfield  {title} {\bibinfo {title} {Calculations of exchange
  bias in thin films with ferromagnetic/antiferromagnetic interfaces},\ }\href
  {https://doi.org/10.1103/PhysRevLett.78.4865} {\bibfield  {journal} {\bibinfo
   {journal} {Phys. Rev. Lett.}\ }\textbf {\bibinfo {volume} {78}},\ \bibinfo
  {pages} {4865} (\bibinfo {year} {1997})}\BibitemShut {NoStop}%
\bibitem [{\citenamefont {Stamps}(2000)}]{stamps2000mechanisms}%
  \BibitemOpen
  \bibfield  {author} {\bibinfo {author} {\bibfnamefont {R.}~\bibnamefont
  {Stamps}},\ }\bibfield  {title} {\bibinfo {title} {Mechanisms for exchange
  bias},\ }\href@noop {} {\bibfield  {journal} {\bibinfo  {journal} {Journal of
  Physics D: Applied Physics}\ }\textbf {\bibinfo {volume} {33}},\ \bibinfo
  {pages} {R247} (\bibinfo {year} {2000})}\BibitemShut {NoStop}%
\bibitem [{\citenamefont {Lu}\ \emph {et~al.}(2000)\citenamefont {Lu},
  \citenamefont {Lai},\ and\ \citenamefont {Chai}}]{lu2000effect}%
  \BibitemOpen
  \bibfield  {author} {\bibinfo {author} {\bibfnamefont {Z.-q.}\ \bibnamefont
  {Lu}}, \bibinfo {author} {\bibfnamefont {W.-y.}\ \bibnamefont {Lai}},\ and\
  \bibinfo {author} {\bibfnamefont {C.-l.}\ \bibnamefont {Chai}},\ }\bibfield
  {title} {\bibinfo {title} {The effect of microstructure and interface
  conditions on the exchange coupling fields of nife/femn},\ }\href@noop {}
  {\bibfield  {journal} {\bibinfo  {journal} {Thin Solid Films}\ }\textbf
  {\bibinfo {volume} {375}},\ \bibinfo {pages} {224} (\bibinfo {year}
  {2000})}\BibitemShut {NoStop}%
\bibitem [{\citenamefont {Kappenberger}\ \emph {et~al.}(2003)\citenamefont
  {Kappenberger}, \citenamefont {Martin}, \citenamefont {Pellmont},
  \citenamefont {Hug}, \citenamefont {Kortright}, \citenamefont {Hellwig},\
  and\ \citenamefont {Fullerton}}]{kappenberger2003direct}%
  \BibitemOpen
  \bibfield  {author} {\bibinfo {author} {\bibfnamefont {P.}~\bibnamefont
  {Kappenberger}}, \bibinfo {author} {\bibfnamefont {S.}~\bibnamefont
  {Martin}}, \bibinfo {author} {\bibfnamefont {Y.}~\bibnamefont {Pellmont}},
  \bibinfo {author} {\bibfnamefont {H.}~\bibnamefont {Hug}}, \bibinfo {author}
  {\bibfnamefont {J.}~\bibnamefont {Kortright}}, \bibinfo {author}
  {\bibfnamefont {O.}~\bibnamefont {Hellwig}},\ and\ \bibinfo {author}
  {\bibfnamefont {E.~E.}\ \bibnamefont {Fullerton}},\ }\bibfield  {title}
  {\bibinfo {title} {Direct imaging and determination of the uncompensated spin
  density in exchange-biased c o o/(c o p t) multilayers},\ }\href@noop {}
  {\bibfield  {journal} {\bibinfo  {journal} {Physical review letters}\
  }\textbf {\bibinfo {volume} {91}},\ \bibinfo {pages} {267202} (\bibinfo
  {year} {2003})}\BibitemShut {NoStop}%
\bibitem [{\citenamefont {Kovalev}\ and\ \citenamefont
  {Pankratova}(2014)}]{kovalev2014field}%
  \BibitemOpen
  \bibfield  {author} {\bibinfo {author} {\bibfnamefont {A.}~\bibnamefont
  {Kovalev}}\ and\ \bibinfo {author} {\bibfnamefont {M.}~\bibnamefont
  {Pankratova}},\ }\bibfield  {title} {\bibinfo {title} {Field dependence of
  magnetization for a thin ferromagnetic film on rough antiferromagnetic
  surface},\ }\href@noop {} {\bibfield  {journal} {\bibinfo  {journal}
  {Superlattices and Microstructures}\ }\textbf {\bibinfo {volume} {73}},\
  \bibinfo {pages} {275} (\bibinfo {year} {2014})}\BibitemShut {NoStop}%
\bibitem [{\citenamefont {Pankratova}\ and\ \citenamefont
  {Kovalev}(2015)}]{pankratova2015model}%
  \BibitemOpen
  \bibfield  {author} {\bibinfo {author} {\bibfnamefont {M.}~\bibnamefont
  {Pankratova}}\ and\ \bibinfo {author} {\bibfnamefont {A.}~\bibnamefont
  {Kovalev}},\ }\bibfield  {title} {\bibinfo {title} {Model of exchange bias in
  a trilayer fm/afm/fm structure},\ }\href@noop {} {\bibfield  {journal}
  {\bibinfo  {journal} {Low Temperature Physics}\ }\textbf {\bibinfo {volume}
  {41}},\ \bibinfo {pages} {838} (\bibinfo {year} {2015})}\BibitemShut
  {NoStop}%
\bibitem [{\citenamefont {Carvalho}\ \emph {et~al.}(2023)\citenamefont
  {Carvalho}, \citenamefont {Miranda}, \citenamefont {Brand{\~a}o},
  \citenamefont {Bergman}, \citenamefont {Cezar}, \citenamefont {Klautau},\
  and\ \citenamefont {Petrilli}}]{carvalho2023correlation}%
  \BibitemOpen
  \bibfield  {author} {\bibinfo {author} {\bibfnamefont {P.~C.}\ \bibnamefont
  {Carvalho}}, \bibinfo {author} {\bibfnamefont {I.~P.}\ \bibnamefont
  {Miranda}}, \bibinfo {author} {\bibfnamefont {J.}~\bibnamefont
  {Brand{\~a}o}}, \bibinfo {author} {\bibfnamefont {A.}~\bibnamefont
  {Bergman}}, \bibinfo {author} {\bibfnamefont {J.~C.}\ \bibnamefont {Cezar}},
  \bibinfo {author} {\bibfnamefont {A.~B.}\ \bibnamefont {Klautau}},\ and\
  \bibinfo {author} {\bibfnamefont {H.~M.}\ \bibnamefont {Petrilli}},\
  }\bibfield  {title} {\bibinfo {title} {Correlation of interface
  interdiffusion and skyrmionic phases},\ }\href@noop {} {\bibfield  {journal}
  {\bibinfo  {journal} {Nano Letters}\ } (\bibinfo {year} {2023})}\BibitemShut
  {NoStop}%
\bibitem [{\citenamefont {Lepadatu}(2019)}]{roughskyrm}%
  \BibitemOpen
  \bibfield  {author} {\bibinfo {author} {\bibfnamefont {S.}~\bibnamefont
  {Lepadatu}},\ }\bibfield  {title} {\bibinfo {title} {Effect of inter-layer
  spin diffusion on skyrmion motion in magnetic multilayers},\ }\href@noop {}
  {\bibfield  {journal} {\bibinfo  {journal} {Scientific Reports}\ }\textbf
  {\bibinfo {volume} {9}},\ \bibinfo {pages} {9592} (\bibinfo {year}
  {2019})}\BibitemShut {NoStop}%
\bibitem [{\citenamefont {Eriksson}\ \emph {et~al.}(2017)\citenamefont
  {Eriksson}, \citenamefont {Bergman}, \citenamefont {Bergqvist},\ and\
  \citenamefont {Hellsvik}}]{eriksson2017atomistic}%
  \BibitemOpen
  \bibfield  {author} {\bibinfo {author} {\bibfnamefont {O.}~\bibnamefont
  {Eriksson}}, \bibinfo {author} {\bibfnamefont {A.}~\bibnamefont {Bergman}},
  \bibinfo {author} {\bibfnamefont {L.}~\bibnamefont {Bergqvist}},\ and\
  \bibinfo {author} {\bibfnamefont {J.}~\bibnamefont {Hellsvik}},\ }\href@noop
  {} {\emph {\bibinfo {title} {Atomistic spin dynamics: foundations and
  applications}}}\ (\bibinfo  {publisher} {Oxford university press},\ \bibinfo
  {year} {2017})\BibitemShut {NoStop}%
\bibitem [{Upp()}]{UppASD}%
  \BibitemOpen
  \href@noop {} {\bibinfo {title} {Uppsala atomistic spin dynamics}},\ \bibinfo
  {note}
  {https://www.physics.uu.se/forskning/materialteori/pagaende-forskning/uppasd/}\BibitemShut
  {NoStop}%
\bibitem [{\citenamefont {Lau}\ and\ \citenamefont
  {Dasgupta}(1989)}]{PhysRevB.39.7212}%
  \BibitemOpen
  \bibfield  {author} {\bibinfo {author} {\bibfnamefont {M.-h.}\ \bibnamefont
  {Lau}}\ and\ \bibinfo {author} {\bibfnamefont {C.}~\bibnamefont {Dasgupta}},\
  }\bibfield  {title} {\bibinfo {title} {Numerical investigation of the role of
  topological defects in the three-dimensional heisenberg transition},\ }\href
  {https://doi.org/10.1103/PhysRevB.39.7212} {\bibfield  {journal} {\bibinfo
  {journal} {Phys. Rev. B}\ }\textbf {\bibinfo {volume} {39}},\ \bibinfo
  {pages} {7212} (\bibinfo {year} {1989})}\BibitemShut {NoStop}%
\bibitem [{\citenamefont {Wang}\ \emph
  {et~al.}(2022{\natexlab{b}})\citenamefont {Wang}, \citenamefont {Wang},
  \citenamefont {Kitamura}, \citenamefont {Hirakata},\ and\ \citenamefont
  {Shimada}}]{wang2022exponential}%
  \BibitemOpen
  \bibfield  {author} {\bibinfo {author} {\bibfnamefont {Y.}~\bibnamefont
  {Wang}}, \bibinfo {author} {\bibfnamefont {J.}~\bibnamefont {Wang}}, \bibinfo
  {author} {\bibfnamefont {T.}~\bibnamefont {Kitamura}}, \bibinfo {author}
  {\bibfnamefont {H.}~\bibnamefont {Hirakata}},\ and\ \bibinfo {author}
  {\bibfnamefont {T.}~\bibnamefont {Shimada}},\ }\bibfield  {title} {\bibinfo
  {title} {Exponential temperature effects on skyrmion-skyrmion interaction},\
  }\href@noop {} {\bibfield  {journal} {\bibinfo  {journal} {Physical Review
  Applied}\ }\textbf {\bibinfo {volume} {18}},\ \bibinfo {pages} {044024}
  (\bibinfo {year} {2022}{\natexlab{b}})}\BibitemShut {NoStop}%
\bibitem [{\citenamefont {Pinna}\ \emph {et~al.}(2020)\citenamefont {Pinna},
  \citenamefont {Bourianoff},\ and\ \citenamefont
  {Everschor-Sitte}}]{PhysRevApplied.14.054020}%
  \BibitemOpen
  \bibfield  {author} {\bibinfo {author} {\bibfnamefont {D.}~\bibnamefont
  {Pinna}}, \bibinfo {author} {\bibfnamefont {G.}~\bibnamefont {Bourianoff}},\
  and\ \bibinfo {author} {\bibfnamefont {K.}~\bibnamefont {Everschor-Sitte}},\
  }\bibfield  {title} {\bibinfo {title} {Reservoir computing with random
  skyrmion textures},\ }\href
  {https://doi.org/10.1103/PhysRevApplied.14.054020} {\bibfield  {journal}
  {\bibinfo  {journal} {Phys. Rev. Appl.}\ }\textbf {\bibinfo {volume} {14}},\
  \bibinfo {pages} {054020} (\bibinfo {year} {2020})}\BibitemShut {NoStop}%
\bibitem [{\citenamefont {Skubic}\ \emph {et~al.}(2008)\citenamefont {Skubic},
  \citenamefont {Hellsvik}, \citenamefont {Nordstr\"om},\ and\ \citenamefont
  {Eriksson}}]{UppASDfirst}%
  \BibitemOpen
  \bibfield  {author} {\bibinfo {author} {\bibfnamefont {B.}~\bibnamefont
  {Skubic}}, \bibinfo {author} {\bibfnamefont {J.}~\bibnamefont {Hellsvik}},
  \bibinfo {author} {\bibfnamefont {L.}~\bibnamefont {Nordstr\"om}},\ and\
  \bibinfo {author} {\bibfnamefont {O.}~\bibnamefont {Eriksson}},\ }\bibfield
  {title} {\bibinfo {title} {A method for atomistic spin dynamics simulations:
  implementation and examples},\ }\href@noop {} {\bibfield  {journal} {\bibinfo
   {journal} {Journal of Physics, Condensed Matter}\ }\textbf {\bibinfo
  {volume} {20}},\ \bibinfo {pages} {315203} (\bibinfo {year}
  {2008})}\BibitemShut {NoStop}%
\bibitem [{\citenamefont {G{\"o}bel}\ \emph {et~al.}(2021)\citenamefont
  {G{\"o}bel}, \citenamefont {Mertig},\ and\ \citenamefont
  {Tretiakov}}]{gobel2021beyond}%
  \BibitemOpen
  \bibfield  {author} {\bibinfo {author} {\bibfnamefont {B.}~\bibnamefont
  {G{\"o}bel}}, \bibinfo {author} {\bibfnamefont {I.}~\bibnamefont {Mertig}},\
  and\ \bibinfo {author} {\bibfnamefont {O.~A.}\ \bibnamefont {Tretiakov}},\
  }\bibfield  {title} {\bibinfo {title} {Beyond skyrmions: Review and
  perspectives of alternative magnetic quasiparticles},\ }\href@noop {}
  {\bibfield  {journal} {\bibinfo  {journal} {Physics Reports}\ }\textbf
  {\bibinfo {volume} {895}},\ \bibinfo {pages} {1} (\bibinfo {year}
  {2021})}\BibitemShut {NoStop}%
\bibitem [{\citenamefont {Reichhardt}\ \emph {et~al.}(2022)\citenamefont
  {Reichhardt}, \citenamefont {Reichhardt},\ and\ \citenamefont {Milo\ifmmode
  \check{s}\else \v{s}\fi{}evi\ifmmode~\acute{c}\else
  \'{c}\fi{}}}]{RevModPhys.94.035005}%
  \BibitemOpen
  \bibfield  {author} {\bibinfo {author} {\bibfnamefont {C.}~\bibnamefont
  {Reichhardt}}, \bibinfo {author} {\bibfnamefont {C.~J.~O.}\ \bibnamefont
  {Reichhardt}},\ and\ \bibinfo {author} {\bibfnamefont {M.~V.}\ \bibnamefont
  {Milo\ifmmode \check{s}\else \v{s}\fi{}evi\ifmmode~\acute{c}\else
  \'{c}\fi{}}},\ }\bibfield  {title} {\bibinfo {title} {Statics and dynamics of
  skyrmions interacting with disorder and nanostructures},\ }\href
  {https://doi.org/10.1103/RevModPhys.94.035005} {\bibfield  {journal}
  {\bibinfo  {journal} {Rev. Mod. Phys.}\ }\textbf {\bibinfo {volume} {94}},\
  \bibinfo {pages} {035005} (\bibinfo {year} {2022})}\BibitemShut {NoStop}%
\bibitem [{\citenamefont {Nagaosa}\ and\ \citenamefont
  {Tokura}(2013)}]{nagaosa2013topological}%
  \BibitemOpen
  \bibfield  {author} {\bibinfo {author} {\bibfnamefont {N.}~\bibnamefont
  {Nagaosa}}\ and\ \bibinfo {author} {\bibfnamefont {Y.}~\bibnamefont
  {Tokura}},\ }\bibfield  {title} {\bibinfo {title} {Topological properties and
  dynamics of magnetic skyrmions},\ }\href@noop {} {\bibfield  {journal}
  {\bibinfo  {journal} {Nature nanotechnology}\ }\textbf {\bibinfo {volume}
  {8}},\ \bibinfo {pages} {899} (\bibinfo {year} {2013})}\BibitemShut {NoStop}%
\end{thebibliography}%

\end{document}